\RequirePackage{amsmath}
\documentclass[runningheads]{llncs}
\usepackage[T1]{fontenc}
\usepackage{graphicx}
\usepackage{amsmath,amssymb,amsfonts}

\usepackage{array}
\usepackage{booktabs}
\usepackage{cite}
\usepackage{color}
\usepackage{colortbl}
\usepackage{comment}
\usepackage{diagbox}
\usepackage{epsfig}
\usepackage{float}
\usepackage{hhline}
\usepackage{lipsum}
\usepackage{listings}
\usepackage{makecell}
\usepackage{multicol}
\usepackage{multirow}
\usepackage{pifont}
\usepackage{subfig}
\usepackage{textcomp}

\usepackage{marvosym}

\usepackage{url}
\usepackage{xspace}
\usepackage{hyperref}
\usepackage[linesnumbered,vlined,ruled]{algorithm2e}
\usepackage{hyphenat}
\PassOptionsToPackage{hyphens}{url}
\newcommand{\blackding}[1]{\ding{\numexpr181+#1\relax}}
\begin{document}
\title{\textsc{BinSimDB}: Benchmark Dataset Construction for \textit{Fine-Grained} Binary Code Similarity Analysis}
\titlerunning{\textsc{BinSimDB}: Benchmark Dataset Construction for Fine-Grained BCSA}
%

\author{Fei Zuo\inst{1}\textsuperscript{(\Letter)} \and Cody Tompkins\inst{1} \and
Qiang Zeng\inst{2} \and Lannan Luo\inst{2} \and \\ 
Yung Ryn Choe\inst{3} \and Junghwan Rhee\inst{1} }
\authorrunning{F. Zuo et al.}
%
\institute{
University of Central Oklahoma, Edmond, OK 73034, USA\\
\email{\{fzuo,ctompkins6,jrhee2\}@uco.edu}
\and
George Mason University, Fairfax, VA 22030, USA\\
\email{\{zeng,lluo4\}@gmu.edu}
\and
Sandia National Laboratories, Livermore, CA 94551, USA\\
\email{yrchoe@sandia.gov}
}

\maketitle              
\begin{abstract}
Binary Code Similarity Analysis (BCSA) has a wide spectrum of applications, including plagiarism detection, vulnerability discovery, and malware analysis, thus drawing significant attention from the security community. However, conventional techniques often face challenges in balancing both accuracy and scalability simultaneously. To overcome these existing problems, a surge of deep learning-based work has been recently proposed. Unfortunately, many researchers still find it extremely difficult to conduct relevant studies or extend existing approaches. First, prior work typically relies on proprietary benchmark without making the entire dataset publicly accessible. Consequently, a large-scale, well-labeled dataset for binary code similarity analysis remains precious and scarce. Moreover, previous work has primarily focused on comparing at the function level, rather than exploring other finer granularities. Therefore, we argue that the lack of a fine-grained dataset for BCSA leaves a critical gap in current research. To address these challenges, we construct a benchmark
dataset for fine-grained binary code similarity analysis called~\textsc{BinSimDB}, which contains equivalent pairs of smaller binary code snippets, such as basic blocks. Specifically, we propose \textsc{BMerge} and \textsc{BPair} algorithms to bridge the discrepancies between two binary code snippets caused by different optimization levels or platforms. Furthermore, we empirically study the properties of our dataset and evaluate its effectiveness for the BCSA research. The experimental results demonstrate that~\textsc{BinSimDB} significantly improves the performance of binary code similarity comparison.

\keywords{Binary analysis \and Benchmark dataset \and Binary code similarity analysis.}
\end{abstract}

\section{Introduction}

The aim of binary code similarity analysis (BCSA) is to determine whether two binary code snippets are equivalent or not. 
A technique for BCSA can be applied to many security-related applications. For example, we take the cross-platform vulnerability search as a motivation problem. Given a patched binary code snippet from one platform, by BCSA we can identify whether there exists a vulnerability due to code reuse in the target binary coming from a different platform. 
However, since the binary code snippets can be generated by using distinct optimization levels or targeting different platforms, the two pieces of code that need to be compared are usually too different to be matched. To address this challenge, a surge of recent work shifts the focus to the popular arsenal of deep learning. For example, graph-based learning methods~\cite{xu2017neural,yu2020order,duan2020deepbindiff,kim2022improving} extract graph information from binary code and use them as the basis for similarity detection because a binary executable is intrinsically associated with some graph representations such as control flow graphs (CFGs). Other methods~\cite{lageman2017bin,ding2019asm2vec,massarelli2019safe,ahn2022practical,wang2022jtrans} consider the assembly code as a language, thus developing NLP-inspired techniques to detect similarity. 

However, high-quality datasets play a significant role in any data-driven application. Thus, the construction of datasets for learning-based security studies has raised awareness among prior researchers. They believe that the unavailability of data is a widespread obstacle for security research~\cite{tian2012identifying,wang2021patchdb,shrestha2023provsec}. This problem is particularly severe in the field of BCSA. For example, after investigating 43 papers in this area, Kim et al.~\cite{kim2022revisiting} found ``\textit{only two of them opened their entire dataset, which makes it difficult to reproduce or extend previous work}''.

Moreover, although a few previous researchers have started making their dataset publicly available for the convenience of subsequent work, they primarily focus on the similarity detection at the function level rather than other finer granularities. Nevertheless, 
the function-level comparison is less useful for the applications 
where subtle discrepancies matter. For example, when patching a piece of vulnerable code, it is usually unnecessary to rewrite the entire function. Therefore, we believe the unavailability of a fine-grained dataset for BCSA requires sufficient attention but has not yet been well resolved. It is worth noting that, as the first deep learning-based approach to detecting binary code similarity at the basic block level, \texttt{InnerEye}~\cite{zuo2019neural} proposed a dataset construction method. However, their method is limited by two drawbacks. 

First, \texttt{InnerEye} extracts binary code directly from the backend of a specific compiler, namely LLVM. However, the binary code obtained from the reverse engineering tool is not exactly identical to the binary code generated by the compiler, which creates a gap in practical application. Second, when matching two equivalent basic block pairs, \texttt{InnerEye} relies on the \textit{annotations} generated by the LLVM facility to annotate every basic block. According to the official manual~\cite{llvm_2023}, the formatted string used by LLVM to identify a basic block and its parent function is ``\textit{hopefully unique}'' but there is no guarantee. Hence, the matching result of equivalent basic blocks is not exactly accurate. Especially when compiling the source code using different optimization levels, it is highly possible that basic blocks from a lower optimization level cannot successfully find an exact match in a new binary obtained by using a higher optimization level. \texttt{InnerEye} suffers from a failure to address this case, thus raising concerns about the data quality. 

Generating basic block-level equivalent pairs is never a trivial task. Previous studies~\cite{kim2022revisiting,massarelli2019safe,lageman2017bin,ahn2022practical} have focused on the similarity between functions, allowing them to establish ground truth using function names. However, unlike functions, basic blocks lack an off-the-shelf unique identifier. Therefore, we propose using source code information to annotate every basic block. Still, establishing a one-to-one mapping among basic blocks across different optimization levels or platforms remains challenging even with source code information. The two main challenges ahead are: \blackding{1} A single line of source code may correspond to multiple basic blocks, and the number of basic blocks originating from the same line of source code may vary across different architectures; \blackding{2} Compiling source code with a higher optimization level may lead to the merging or reorganization of original basic blocks generated at a lower optimization level due to compiler optimization behavior. To address these challenges in dataset construction, we have developed~\textsc{BMerge} and \textsc{BPair} algorithm. We will shed light on our observations using a concrete motivation example and also the proposed solutions in Section~\ref{sec:method}.

The key contributions of our work include:

\begin{itemize}

\item We construct a fine-grained dataset \textsc{BinSimDB}\footnote[1]{We share our dataset with the cybersecurity community in the following link: \url{https://uco-cyber.github.io/research/\#binsimdb}.}, 
which consists of 4,426,258 equivalent assembly code pairs, for facilitating BCSA studies. To the best of our knowledge, we are the ﬁrst to release a diverse set of fine-grained equivalent binary code pairs at this scale.

\item Not only the comprehensive benchmark for BCSA, we also make our automated scripts publicly accessible, so that future academics are able to easily reproduce or extend the dataset for various research purposes. 

\item We construct the reliable ground truth by adopting source code information. In particular, we propose \textsc{BMerge} and \textsc{BPair} algorithm to handle the issues raised in the equivalent binary code matching at a fine granularity.

\item We empirically investigate the properties of \textsc{BinSimDB}, and evaluate its effectiveness for the BCSA research. The experimental results demonstrate that our dataset can greatly help to improve the performance of binary code similarity comparison.

\end{itemize}


\section{Background}\label{sec:bck}


\subsection{Binary Code Similarity Analysis}

Binary code similarity analysis (BCSA) is the process of comparing two or more binary code snippets to determine how similar they are. The goal of this analysis is to identify potential code reuse or plagiarism between different software programs, as well as to discover potential security vulnerabilities or malware. Note that in this paper, we will use the terms `binary code' and `assembly code' interchangeably, unless otherwise specified.

\begin{figure}[!th]
\centering
  \includegraphics[width=0.54\textwidth]{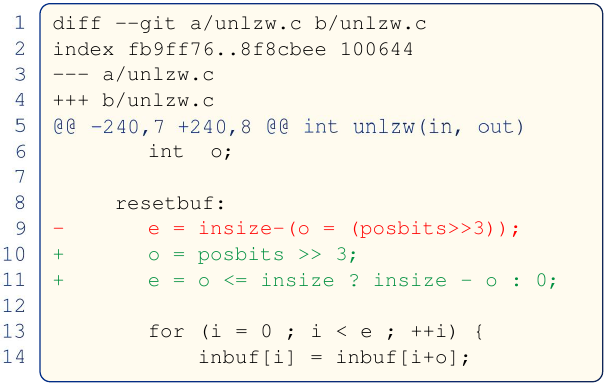}
  \caption{A patch in \texttt{gzip} for CVE-2010-0001.}
  \label{fig:cve_gzip}
\end{figure}

\begin{figure} [!th]
\centering
\subfloat[Binary code in x86-64]{\includegraphics[scale=0.34]{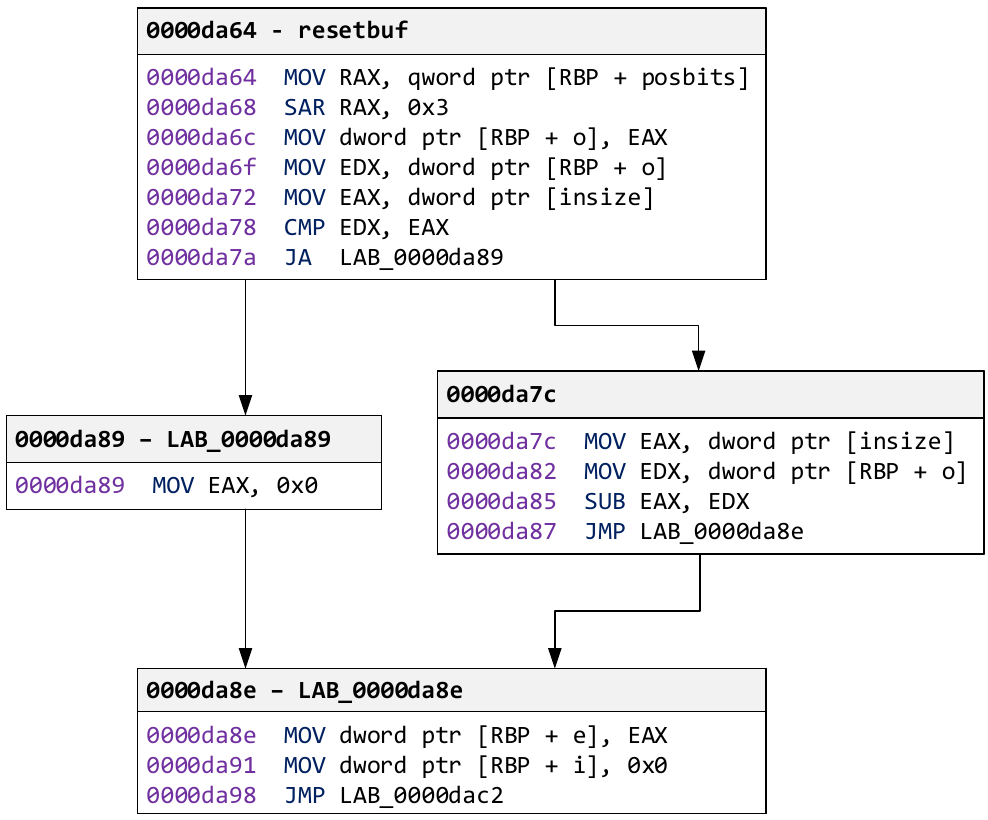}}\quad
\subfloat[Binary code in AArch64]{\includegraphics[scale=0.34]{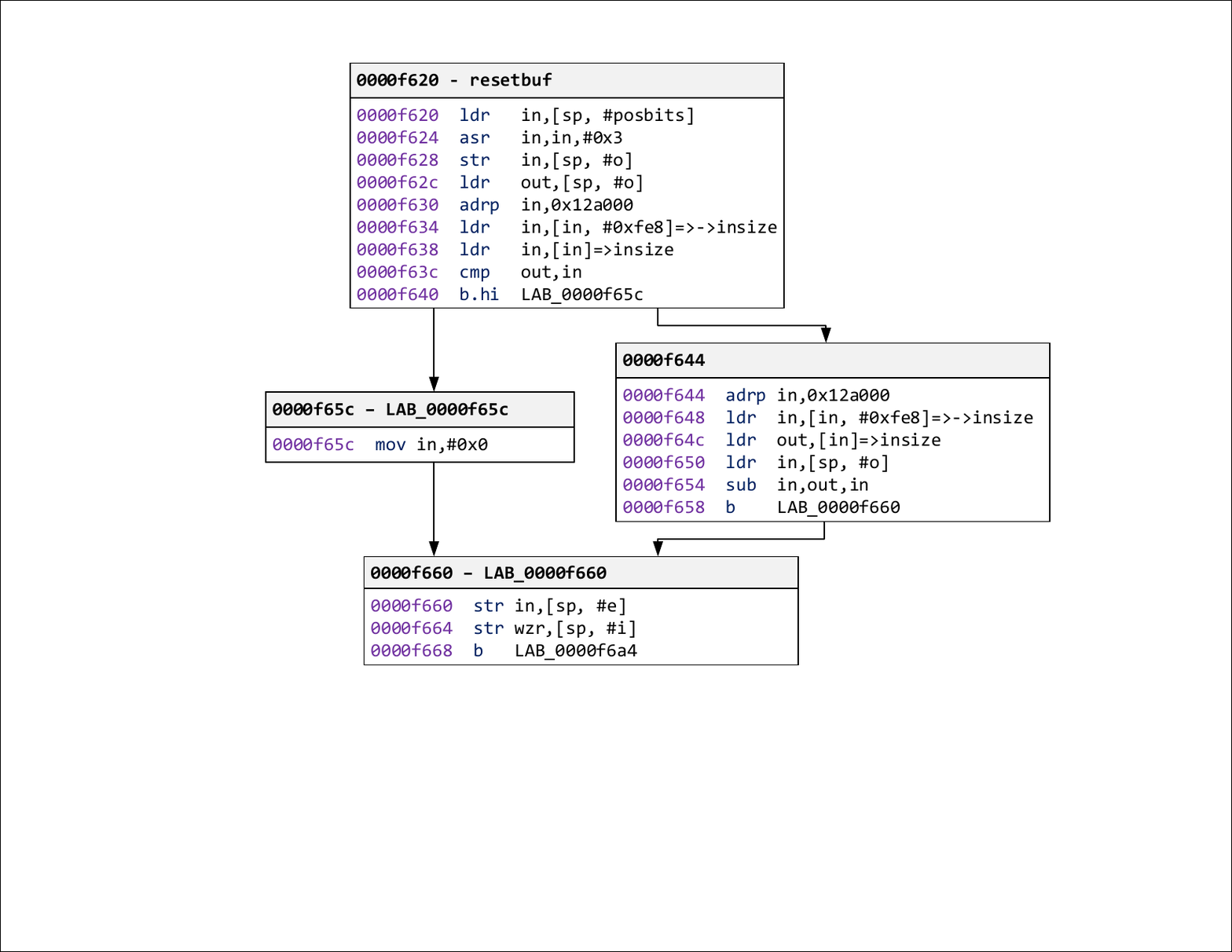}}
\caption{A pair of equivalent binary code snippets from the same source code.}\label{fig:patch}
\end{figure}

We regard two binary code snippets as similar if they come from the same piece of source code. For example, Figure~\ref{fig:cve_gzip} shows a source code patch for CVE-2010-0001. In detail, there is an integer underflow vulnerability in the \texttt{unlzw} function of \texttt{gzip} before 1.4 on 64-bit platforms, thus possibly causing a denial of service or allowing remote attackers to execute arbitrary code. We compiled the above patch code for two different platforms, x86-64 and AArch64, resulting in two equivalent binary code snippets, as shown in Figure~\ref{fig:patch}(a) and Figure~\ref{fig:patch}(b), respectively.
For the sake of presentation, we particularly use the optimization level \texttt{O0} when compiling the patched code. Though control flow graphs (CFGs) across the two platforms remain similar, the binary code looks quite different because of distinct instruction set architectures (ISAs) and registers. Furthermore, there are 75 basic blocks in the \texttt{unlzw} function overall, while code changes caused by the patch only take up approximately 5.3\% of the entire function. Hence, this example showcases that considering fine-grained discrepancy in some practical applications is of great necessity as well.

\subsection{Toolchain}
It is noteworthy that the third-party applications utilized in this study are non-proprietary, which provides sufficient flexibility for the academic community to re-use all the research resources. In detail, we develop our system using Python in Ubuntu 22.04. For the software reverse engineering tool, we choose the free and open sourced framework \texttt{Ghidra}\footnote[2]{\url{https://ghidra-sre.org/}}, which is developed by the National Security Agency (NSA) of the United States. \texttt{Ghidra} is capable of analyzing compiled code on any platform, whether Linux, Windows, or macOS. On top of that, \texttt{Ghidra} also enables users to perform automated analysis with scripts in Python via \texttt{Ghidrathon} extension. 
Today, \texttt{Ghidra} has been widely used by the security community~\cite{pang2021sok} and is also regarded as well-matched in strength with 
its expensive competitor \texttt{IDA Pro}\footnote[3]{\url{https://hex-rays.com/ida-pro/}}~\cite{votipka2021investigation}. 

Furthermore, the GNU binary utility \texttt{addr2line}\footnote[4]{\url{https://www.gnu.org/software/binutils/}} is adopted to translate hexadecimal addresses in a executable into source code file names and line numbers. Given an address in an executable or an offset in a section of a relocatable object, it uses the debugging information to figure out which file name and line number, in the source code, are associated with it. By this means, we can annotate every basic block using source code information. As a result, basic block splitting and combination, or function inline will not be a problem for building a self-evident connection between two assembly code snippets even if they are from different architectures or optimization levels. 
Besides, \texttt{llvm-addr2line} developed by the LLVM project can be used as a drop-in replacement for GNU’s \texttt{addr2line}. The two utilities are interchangeable in this project.

\section{Methodology}\label{sec:method}

\subsection{Key Observations}

\begin{figure}[!htb]
  \centering
  \includegraphics[width=0.5\textwidth]{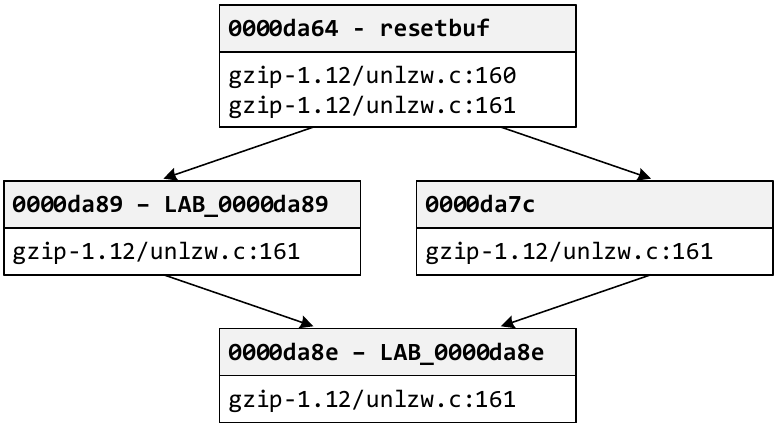}
  \caption{Basic block annotation with corresponding source file and line numbers.}
  \label{fig:addr2line}
\end{figure}

To construct the ground truth using source code information, we traverse all addresses for a given basic block $i$ and leverage \texttt{addr2line} to obtain a label set $\mathcal{A}_i$, where each element corresponds to a source file and line number associated with the basic block. 
For example, we annotate every basic block in Figure~\ref{fig:patch}(b), and the result is shown in Figure~\ref{fig:addr2line}. However, we observe that \emph{a single line of source code may correspond to multiple basic blocks}. Consequently, there may be two different basic blocks $i$ and $j$ in the same function, such that $\mathcal{A}_i \cap \mathcal{A}_j \neq \emptyset$. This inevitably leads to confusion when pairing two basic blocks across different optimization levels or platforms. To ensure the uniqueness of the annotation for a binary code unit under a given setting, we propose using the \textsc{BMerge} Algorithm (shown in Algorithm 1) to handle those overlapping basic blocks.

\begin{figure*}[!htb]
  \centering
  \includegraphics[width=1\textwidth]{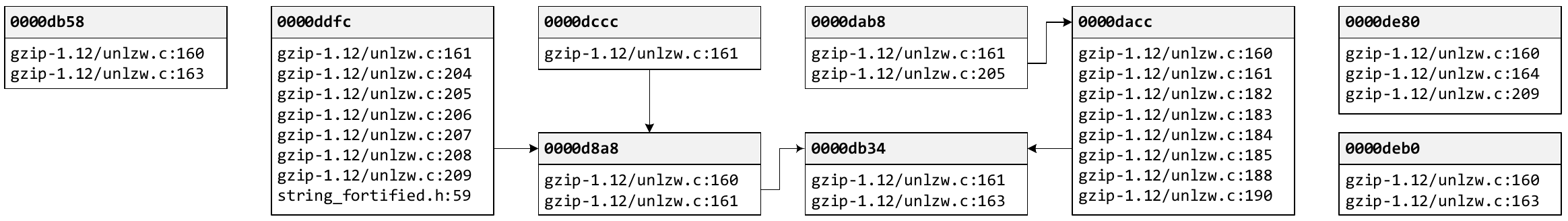}
  \caption{Binary code in AArch64 obtained by using \texttt{O3} as the optimization level.}
  \label{fig:cfg3}
\end{figure*}

Our second observation is, due to the nature of compilers, \emph{multiple basic blocks may be recombined or spread across other basic blocks} when using higher optimization levels. For example, when compiling the patched code in Figure~\ref{fig:patch}(a) using \texttt{O3} on the AArch64 platform, no specific basic blocks obtained are purely derived from the patched code. Instead, the original four basic blocks in Figure~\ref{fig:patch}(c) are recombined and spread across another nine basic blocks, as shown in Figure~\ref{fig:cfg3}.
We can also see some isolated basic blocks. Note that they need to connect to the CFG through some other basic blocks rather than being directly linked. The basic blocks used to connect all the nodes in Figure~\ref{fig:cfg3} are not depicted if they are entirely generated from source code beyond the patch.
This phenomenon introduces ambiguity when paring two basic blocks across different optimization levels. To this end, we propose the \textsc{BPair} Algorithm (as shown in Algorithm 2) to match equivalent assembly code units.

\subsection{Dataset Construction Approach Details}\label{sec:details}

\begin{figure*}[!htb]
  \centering
  \includegraphics[width=1\textwidth]{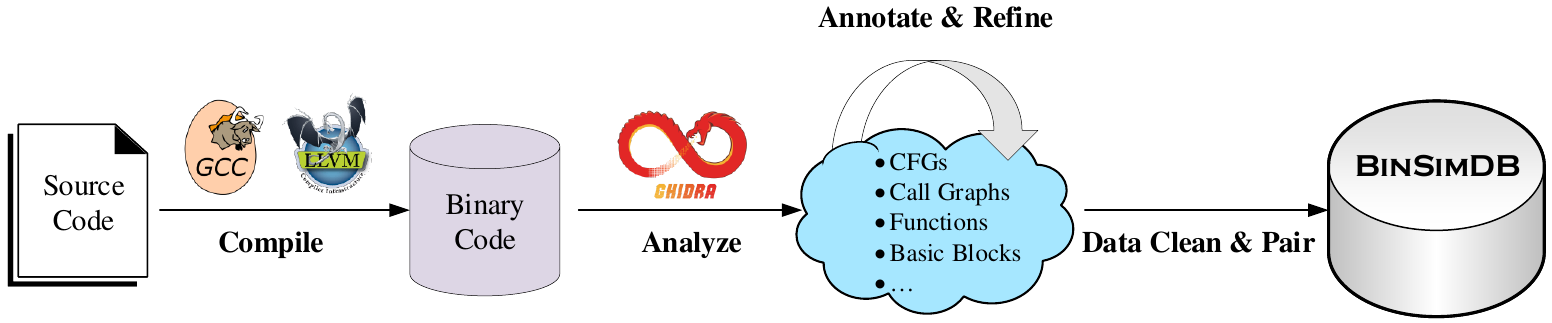}
  \caption{System overview.}
  \label{fig:overview}
\end{figure*}

Figure~\ref{fig:overview} illustrates the construction method of \textsc{BinSimDB}, where all the five steps can be automated by scripts.
First, we compile all the binaries with debugging information using the \texttt{-g} option. Specifically, source files are complied with two representative compilers (GCC and Clang) across various optimization levels (\texttt{O0}, \texttt{O1}, \texttt{O2}, and \texttt{O3}). Four popular ISAs are considered, including x86, x86-64, ARM32, and AArch64. Users can easily extend the script through configuring it to cover additional ISAs.

\begin{algorithm} 
\caption{\textsc{BMerge} Algorithm} 
\KwIn{A set of basic blocks, denoted by $\mathcal{S}$, wherein a basic block $i$ is labeled with a set $\mathcal{A}_i$, and $addr_{i}$ represents the address of $i$.} 
\KwOut{The refined basic blocks set $\mathcal{S}$.} 
\ForEach {$i, j\in \mathcal{S}$, where $i\neq j$,} 
{
\If{$\mathcal{A}_i= \mathcal{A}_j$}
{
\If{$addr_{i} < addr_{j}$}
{
Update $i$ by merging $j$ into $i$\\
$\mathcal{S} \leftarrow \mathcal{S} - \{j\}$\\
}
\Else
{
Update $j$ by merging $i$ into $j$\\
$\mathcal{S} \leftarrow \mathcal{S} - \{i\}$\\
}
}
\ElseIf{$\mathcal{A}_j\subset \mathcal{A}_i$}
{
Update $i$ by merging $j$ into $i$\\
$\mathcal{S} \leftarrow \mathcal{S} - \{j\}$\\
}
\ElseIf{$\mathcal{A}_i\subset \mathcal{A}_j$}
{
Update $j$ by merging $i$ into $j$\\
$\mathcal{S} \leftarrow \mathcal{S}-\{i\}$\\
}
}
\textbf{return} $S$
\end{algorithm}

In the second step, we utilize the facilities provided by \texttt{Ghidra} to analyze the compiled binaries. This allows us to collect static analysis results such as binary functions, basic blocks, control flow graphs (CFGs), and call graphs. Notably, we preliminarily sanitize the dataset to exclude external functions that lack actual function bodies. However, for functions from widely-used third-party libraries such as \texttt{glibc}, we maintain a dictionary $\mathcal{D}$ to record them. Modern reverse engineering applications like \texttt{Ghidra} and \texttt{IDA Pro} have developed techniques for identifying library functions, so we do not focus on this aspect in our work.

Next, we utilize \texttt{addr2line} to annotate every basic block with the corresponding information of source files and line numbers. Specifically, functions generated by compilers rather than the application itself are discarded. This can be easily achieved by referencing the source code information obtained from \texttt{addr2line}. More concretely, we discard a basic block $i$ if $\mathcal{A}_i=\emptyset$, where $\mathcal{A}_i$ is the label set consisting of source file names and line numbers. After that, we further refine our dataset via the \textsc{BMerge} Algorithm (as shown in Algorithm 1). For any two basic blocks $i$ and $j$, they need to be merged into one block as long as their label sets $\mathcal{A}_i$ and $\mathcal{A}_j$ can fulfill any following condition: \blackding{1}$\mathcal{A}_i$ and $\mathcal{A}_j$ are identical (Line 2$\sim$8); \blackding{2}$\mathcal{A}_j$ is the subset of $\mathcal{A}_i$ (Line 9$\sim$11); \blackding{3}$\mathcal{A}_i$ is the subset of $\mathcal{A}_j$ (Line 12$\sim$14). For example, by applying this algorithm, the four basic blocks shown in Figure~\ref{fig:patch}(b) can be integrated into a new block. This resulting block exactly reflects the patched code in Figure~\ref{fig:patch} and does not overlap with any other basic blocks. The \textsc{BMerge} Algorithm does not need to be applied in the following two possible cases: 1) there is only one basic block in a given function; 2) every basic block is obtained from a few lines of source code that are not simultaneously used to generate any other basic blocks.

To control the vocabulary size and preserve the semantics of instructions, many previous studies~\cite{gao2021lightweight,zuo2019neural,duan2020deepbindiff,li2021palmtree} perform normalization on instructions. Similarly, we preprocess the instructions in the dataset according to the following empirical rules: \blackding{1} For numeric constants, according to the sign of the value, we replace a numeric constant with \texttt{<POSITIVE>}, \texttt{<NEGATIVE>} or \texttt{<ZERO>}; \blackding{2} For function calls, if the library function can be identified (i.e. the function name is collected by the dictionary $\mathcal{D}$), we preserve the instruction as its original form. Otherwise, the function names are uniformly replaced with \texttt{<FOO>}; \blackding{3} The memory addresses, such as the local destination of a jump instruction, are replaced by \texttt{<ADDRESS>}; \blackding{4} Finally, we substitute the token \texttt{<STRING>} for all string literals.

\begin{algorithm} 
\caption{\textsc{BPair} Algorithm} 
\KwIn{Two sets of basic blocks, denoted by $\mathcal{U}$ and $\mathcal{V}$, where a basic block $i$ is labeled with a set $\mathcal{A}_i$, and $addr_{i}$ represents the address of $i$.}
\KwOut{A set $\mathcal{M}$ consisting of equivalent basic block pairs.}

\SetKwFunction{BM}{Merge}

\SetKwProg{myalg}{Function}{}{}
\myalg{\BM{$p, q$}}{
\If{$addr_{p} < addr_{q}$}
{
Update $p$ by merging $q$ into $p$\\
\KwRet $p$\\
}
\Else
{
Update $q$ by merging $p$ into $q$\\
\KwRet $q$\\
}
}

Initialize a bipartite graph $\mathcal{G}=(\mathcal{U}, \mathcal{V}, \mathcal{E})$, where $\mathcal{E}=\emptyset$\\
\ForEach {$u\in \mathcal{U}, v\in \mathcal{V}$,}
{
  \If{$\mathcal{A}_u\cap \mathcal{A}_v\neq \emptyset$}
  {
      $\mathcal{E} \leftarrow \mathcal{E} \cup \{\langle u,v \rangle$\} \\
  }
}
  
\ForEach {connected sub-graph $\mathcal{C} \subset \mathcal{G}$}
{
  Pick any basic block $i$ from $\mathcal{C}$, where $i \in \mathcal{U}$ \\
  Pick any basic block $j$ from $\mathcal{C}$, where $j \in \mathcal{V}$ \\
  %
  \ForEach{$k \in \mathcal{C}-\{i,j\}$}{
  
  \If{$k \in \mathcal{U}$}
  {
   $i \leftarrow $ \BM($k$, $i$)\\
  }
  \ElseIf{$k \in \mathcal{V}$}
  {
   $j \leftarrow $ \BM($k$, $j$)\\
  }
  }
  $\mathcal{M} \leftarrow \mathcal{M} \cup \{(i,j)\}$\\
}
\KwRet{$\mathcal{M}$}
\end{algorithm}

\begin{figure*}[!th]
\centering
  \includegraphics[width=0.82\textwidth]{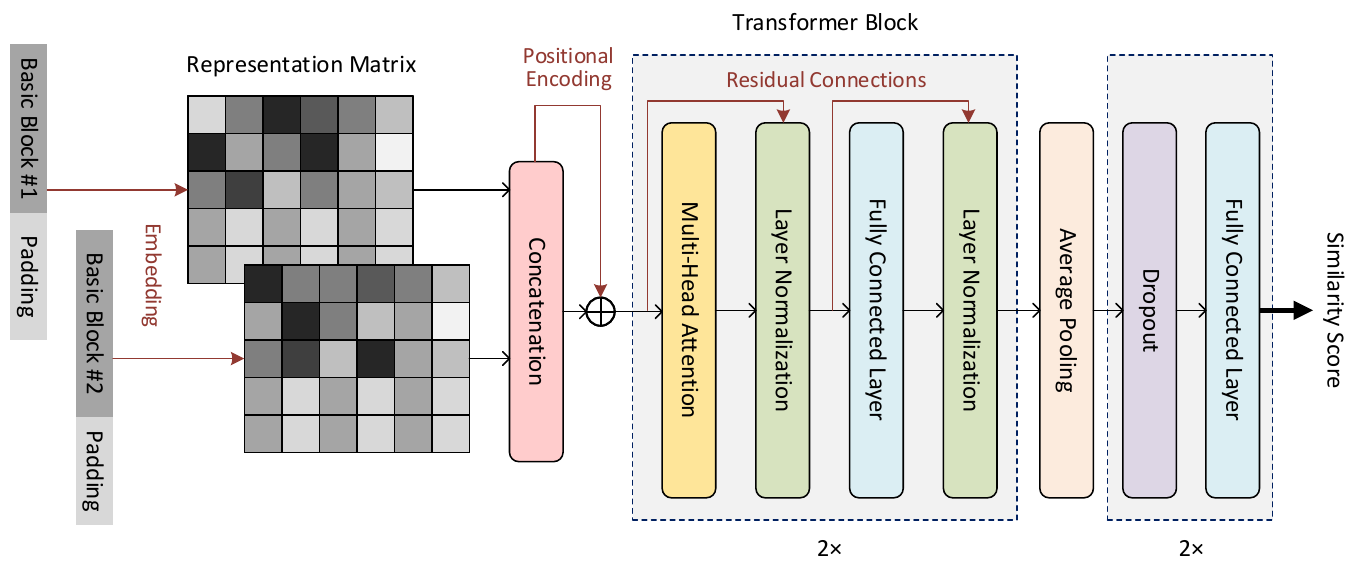}
  \caption{Architecture of the proposed binary code similarity detector.}
  \label{fig:classifier}
\end{figure*}

At the final stage, we generate equivalent pairs by adopting the \textsc{BPair} Algorithm (as shown in Algorithm 2). The intuition behind this algorithm is to transform the problem of matching equivalent basic blocks into a graph problem. 
First, the source code of a function can be compiled under different platforms or using different optimization levels, resulting in two sets of basic blocks denoted by $\mathcal{U}$ and $\mathcal{V}$. 
Based on this, we can build a disconnected bipartite graph $\mathcal{G}$ with partitions $\mathcal{U}$ and $\mathcal{V}$ (Line 8), where the basic blocks $u\in \mathcal{U}$, $v\in \mathcal{V}$ are considered as vertices. 
Then, we create edges continuously to connect vertices (i.e. basic blocks) in $\mathcal{U}$ to those in $\mathcal{V}$, according to the ground truth regarding basic blocks (Line 9$\sim$11). In other words, for basic blocks 
$u\in \mathcal{U}$ and $v\in \mathcal{V}$ that are from two different architectures or optimization levels, if their source code overlaps, i.e., the intersection of $\mathcal{A}_u$ and $\mathcal{A}_v$ is not empty (Line 10), then an edge is established between the two vertices $u$ and $v$ (Line 11). After traversing every connected sub-graph of $\mathcal{G}$ based on a disjoint set, we can generate a set of equivalent pairs $\mathcal{M}$ (Line 12$\sim$20). Additionally, all the duplicate pairs are removed from this set to maintain the dataset quality. 


\subsection{Similarity Detection Model}\label{sec:classifier}

The success of OpenAI's ChatGPT~\cite{brown2020language} has sparked significant interest in the academic and industry sectors regarding Large Language Models (LLMs). Herein, the transformer architecture 
plays a crucial role~\cite{vaswani2017attention}, thus was considered as the fundamental building blocks of LLMs.
To demonstrate the benefits of ~\textsc{BinSimDB} in supporting the future BCSA research, we introduce a Transformer-based binary code similarity detector, as depicted in Figure~\ref{fig:classifier}. 

At first, a pair of assembly code snippets under different architectures are represented by two embedding matrices. The concatenated input of the two matrices and the corresponding positional encoding are provided to a Transformer-based model. The position encoding aims to capture the order information of each instruction within a code snippet. 
These generated positional embeddings are added to the concatenated matrix, then sent together to the subsequent Transformer blocks. The embedding input, in the form of three learnable weight matrices including queries $\mathbf{Q}$, keys $\mathbf{K}$ of dimension $d_k$, and values $\mathbf{V}$ of dimension $d_v$, is passed through a scaled dot-product attention. Formula (1) clearly describes the performance of an attention model.
\begin{equation}
    Attention(\mathbf{Q},\mathbf{K},\mathbf{V}) = Softmax\left(\frac{\mathbf{Q}\mathbf{K}^{T}}{\sqrt{d_k}}\right)\mathbf{V}
\end{equation}
Then, the results are concatenated through a multi-head attention, where each result of the parallel computations of attention is called a head.
\begin{equation}
MultiHead(\mathbf{Q},\mathbf{K},\mathbf{V}) = [head_1, \cdots, head_h]\mathbf{W}^O
\end{equation}
where $head_i=Attention(\mathbf{QW}_i^Q,\mathbf{KW}_i^K,\mathbf{VW}_i^V)$, and every $\mathbf{W}$ represents a learnable parameter matrix. In addition, we also place the layer normalization between the residual blocks inside a transformer block as Formula (3) and (4) show. 
\begin{align}
Output &= LayerNorm(Input + MultiHead(Input)) \\
Output &= LayerNorm(Output + FFN(Output))
\end{align}
where $FFN$ represents the fully connected feed forward layer. 

In a nutshell, we use the Transformer blocks as an encoder that can analyse the assembly code pair and generates a series of hidden states that capture the semantics and global context of the inputs. The subsequent linear layers finalize the prediction with outputting a similarity score. When training the model, we utilize Adam optimizer~\cite{kingma2015adam} with a sparse categorical cross-entropy loss function to improve the learning rate.

It is necessary to emphasize that the focus of this paper is on the construction of the dataset. Consequently, the the binary code similarity detector presented herein serves solely as a means to demonstrate \textsc{BinSimDB}'s usage in reality. Based on this, we will showcase the advantage of our dataset over prior competitors in Section~\ref{sec:comp}.

\section{Evaluation}\label{sec:eval}

Employing the methodology introduced in Section~\ref{sec:method}, we build a 
fine-grained dataset~\textsc{BinSimDB} as a benchmark for BCSA study. In this section, empirical studies are conducted to investigate the dataset properties. Also, the extensive evaluations showcase the 
strength of~\textsc{BinSimDB} for the related research.

\subsection{Dataset Properties and Composition}

\begin{figure}[!thb]
\centering
\subfloat[x86-64]{\includegraphics[scale=0.34]{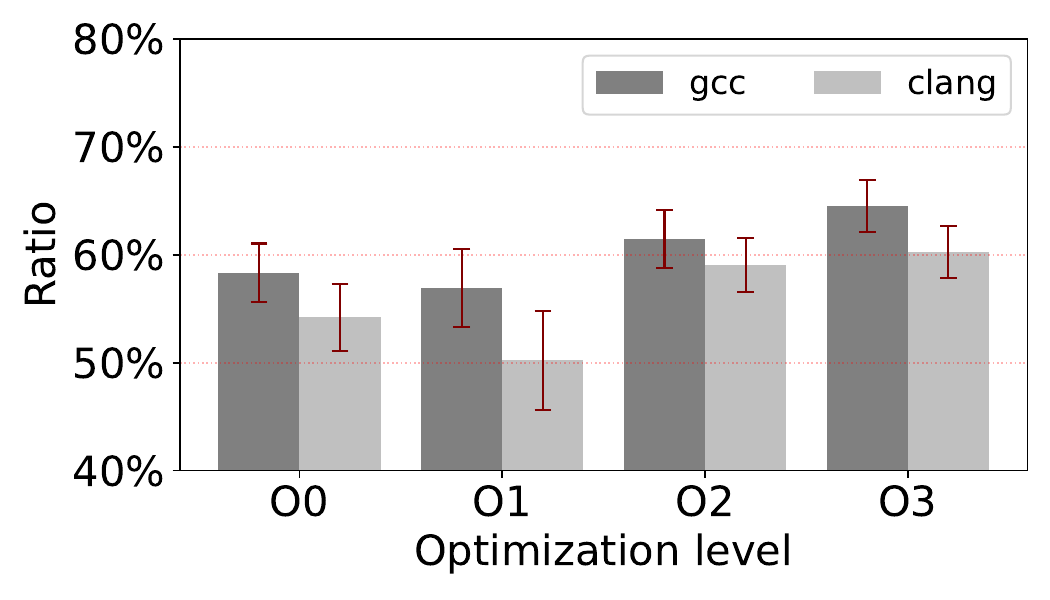}} 
\subfloat[x86]{\includegraphics[scale=0.34]{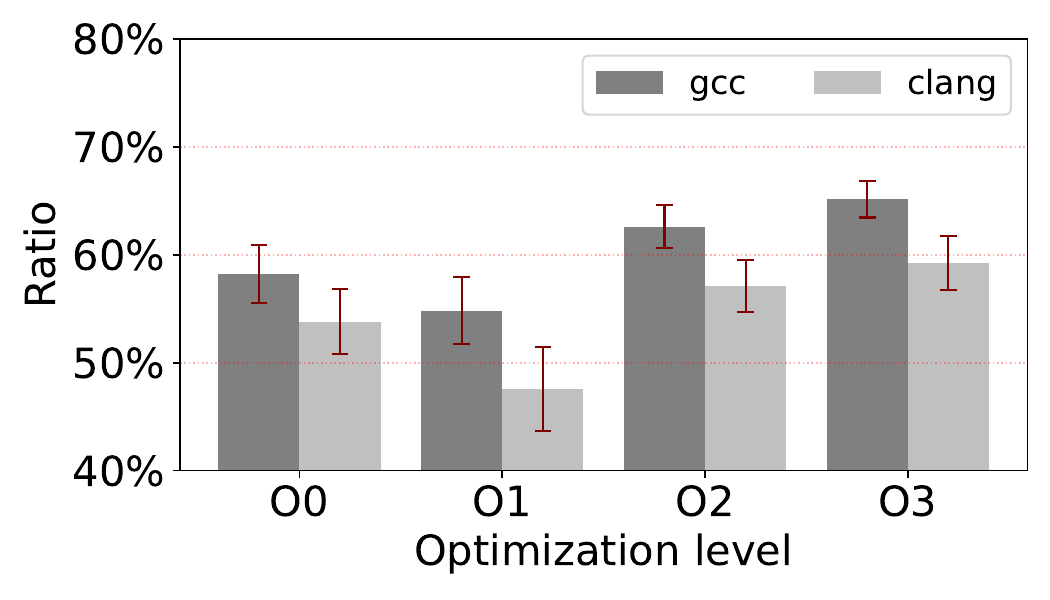}}\\
\subfloat[AArch64]{\includegraphics[scale=0.34]{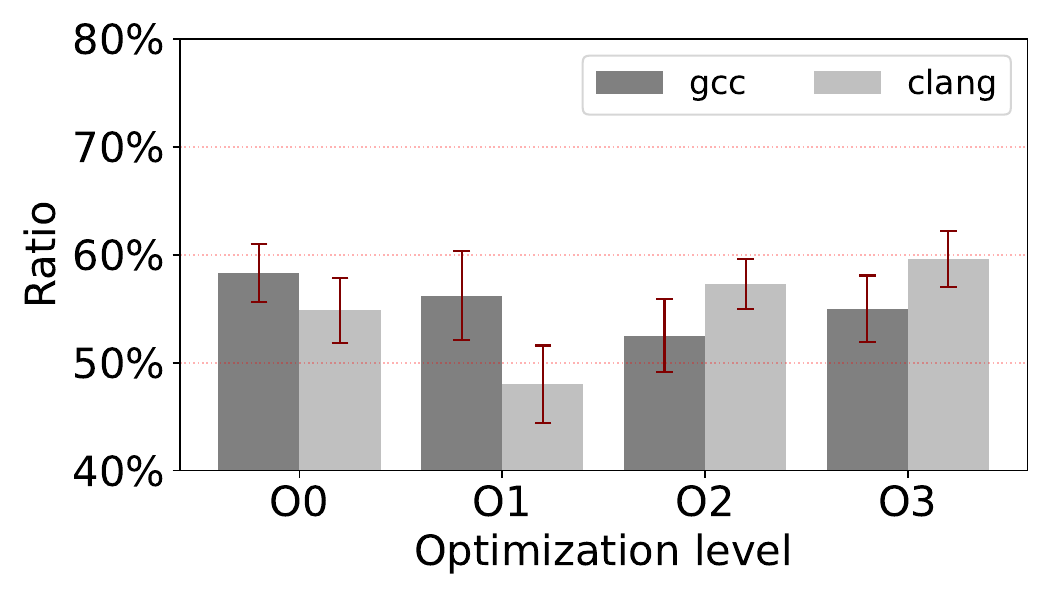}}
\subfloat[ARM32]{\includegraphics[scale=0.34]{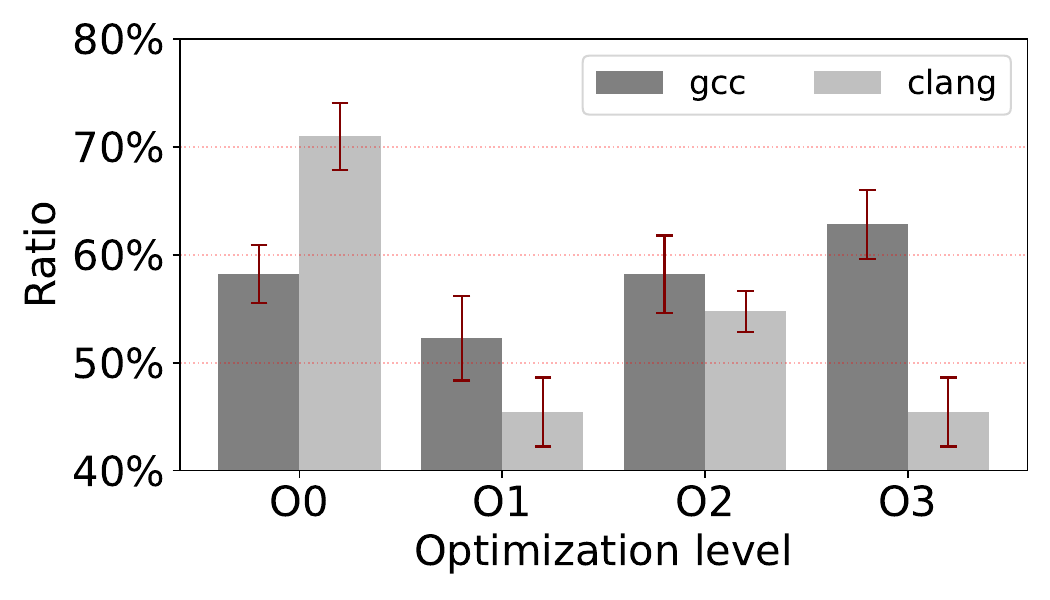}}
\caption{The average ratio of the functions that need to be processed by \textsc{BMerge} to all functions, where the error bars indicate the 95\% confidence intervals. }\label{bmerg_stat1}\vspace{-16pt}
\end{figure}

\noindent\textbf{Diversity and Scalability.} To generate binary samples, we collect source code of 30 binaries from 8 GNU software projects~\cite{gnu_2023}, i.e., \texttt{binutils},  \texttt{datamash}, \texttt{findutils}, \texttt{grep}, \texttt{gzip}, \texttt{macchanger}, \texttt{tar}, and \texttt{which}. They are all real programs and widely deployed in the current software ecosystem. More importantly, because their source code is publicly accessible, GNU packages have became a very popular research resource for BCSA~\cite{lageman2017bin, zuo2019neural, ahn2022practical, wang2022jtrans}. Our comprehensive BCSA benchmark \textsc{BinSimDB} involves 980,251 functions across 32 distinct combinations of compilers, optimization levels, and target platforms. 
These functions will be used to further generate equivalent assembly code pairs. 
More specifically, we include binaries compiled for four different ISAs such as x86, x86-64, ARM32, and AArch64. Two representative compilers, i.e., GCC and Clang, are involved. We also consider four optimization levels from O0 to O3. On top of that, we develop automated scripts to compile the collected source code and disassemble the resulting binaries for all designated architectures and optimization levels. Therefore, other researchers can extend the existing dataset with little effort, or customize their dataset generation towards diverse application scenarios such as IoT applications analysis.

\noindent\textbf{Fine Granularity.} Unlike other existing work~\cite{zuo2019neural}, we do not drop any assembly code obtained from the disassembler, so our dataset provides a good coverage. Not only that, owing to our proposed algorithms, we can generate semantically equivalent assembly code pairs at a finer granularity. 

\begin{figure}[!thb]
\centering
\subfloat[x86-64]{\includegraphics[scale=0.34]{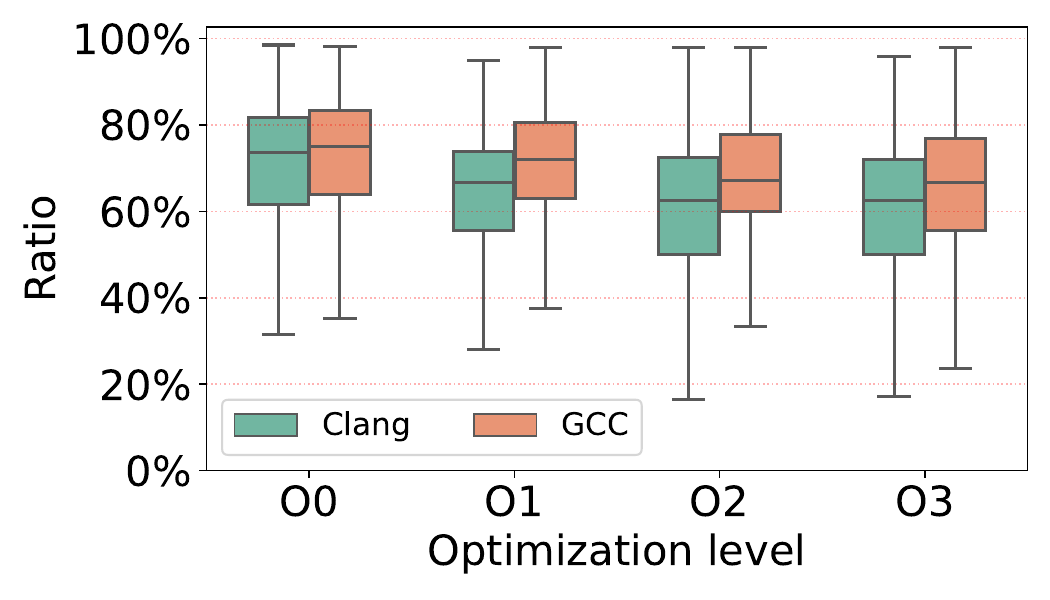}}
\subfloat[x86]{\includegraphics[scale=0.34]{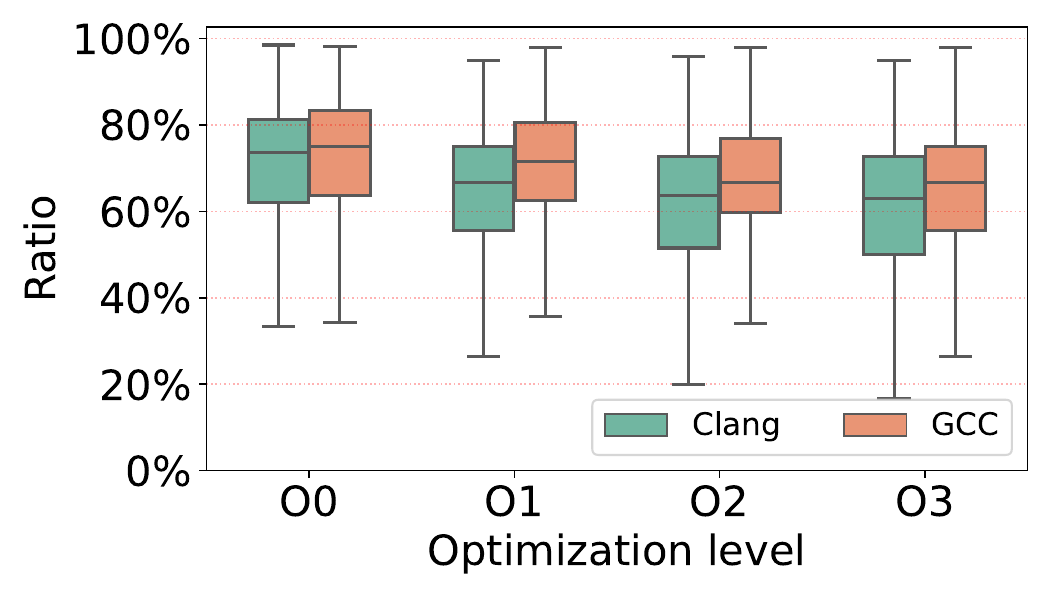}}\\
\subfloat[AArch64]{\includegraphics[scale=0.34]{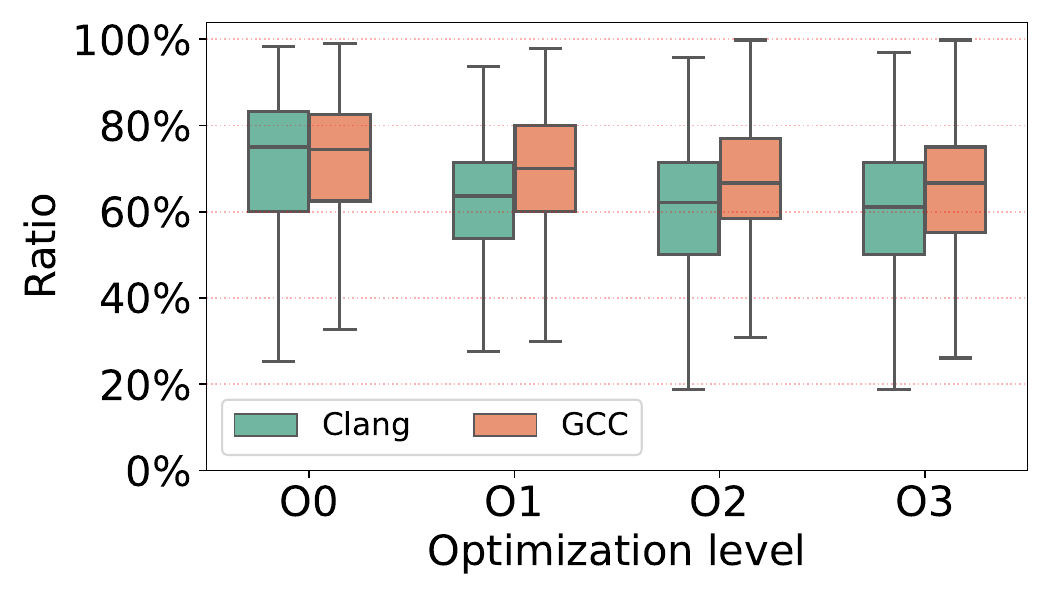}}
\subfloat[ARM32]{\includegraphics[scale=0.34]{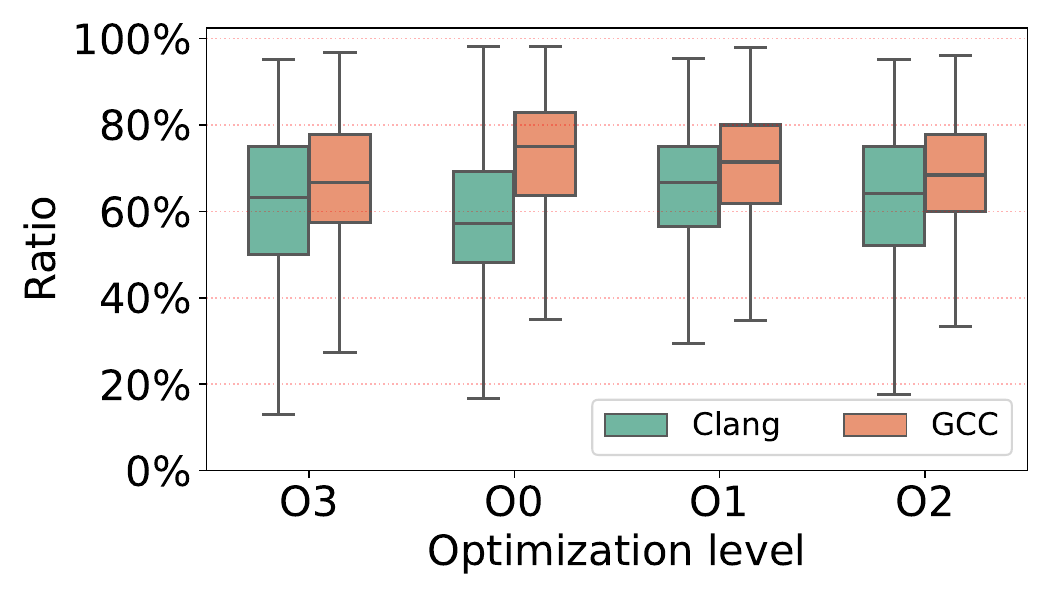}}
\caption{The changes in the number of basic blocks after applying \textsc{BMerge} (Ratio = $\frac{\#\ of\ resulting\ basic\ blocks}{\#\ of\ original\ basic\ blocks}$). }\label{bmerg_stat2}
\end{figure}

It is worth noting that not all functions need to be handled by the \textsc{BMerge}. If the source code snippets used to generate basic blocks are mutually exclusive, it is surely unnecessary to merge any such basic blocks. 
To quantitatively study the cases where a one-to-one mapping cannot be established between basic blocks and source code, we conducted program-level statistics on the proportion of the functions containing such basic blocks among all the functions. The results are shown in Figure~\ref{bmerg_stat1}. As is evident, the ratio of the functions that need to be processed by the \textsc{BMerge} to all functions is around 45\% $\sim$ 65\%, by average. On the ARM32 platform, when using \texttt{O3} as the optimization level, this ratio even reaches 70.9\%. Therefore, it can be concluded that the phenomenon of lacking one-to-one mapping between basic blocks and source code is widespread in practice.

We also count the number of functions, on which we conduct \textsc{BMerge}, and show the statistics in Table~\ref{tab_bmerge}. Based on this, we investigate the impact of applying the \textsc{BMerge} algorithm on the number of basic blocks, as Figure~\ref{bmerg_stat2} shows. It can be seen that the distribution of changes in the number of basic blocks is diverse in terms of different architectures and optimization levels. However, the majority of functions are likely to see a reduction in the number of their basic blocks to approximately 80\% or below of the original count.

\begin{table}[!htbp]
\vspace{-2.0em}
\centering
\caption{The number of functions that need to be processed by \textsc{BMerge}. These functions are leveraged to investigate the impact of our proposed algorithm.}
\renewcommand\arraystretch{1.3}
\begin{tabular}{|l|l|llll|}
\hline
\multicolumn{1}{|c|}{\multirow{2}{*}{\textbf{ISAs}}} & \multicolumn{1}{c|}{\multirow{2}{*}{\textbf{Compilers}}} & \multicolumn{4}{c|}{\textbf{Optimization levels}}                                                                                         \\ \cline{3-6} 
\multicolumn{1}{|c|}{}                               & \multicolumn{1}{c|}{}                                    & \multicolumn{1}{c|}{\textbf{O0}} & \multicolumn{1}{c|}{\textbf{O1}} & \multicolumn{1}{c|}{\textbf{O2}} & \multicolumn{1}{c|}{\textbf{O3}} \\ \hline
\multicolumn{1}{|l|}{\multirow{2}{*}{x86}}           & GCC                & \multicolumn{1}{l|}{19,765}            & \multicolumn{1}{l|}{15,492}            & \multicolumn{1}{l|}{16,909}            &      16,358        \\ \cline{2-6} 
\multicolumn{1}{|c|}{}                               & Clang                                                    & \multicolumn{1}{l|}{18,565}            & \multicolumn{1}{l|}{17,571}            & \multicolumn{1}{l|}{13,529}            &  14,067    \\ \hline
\multirow{2}{*}{x86-64}                              & GCC                                                      & \multicolumn{1}{l|}{19,784}      & \multicolumn{1}{l|}{16,075}      & \multicolumn{1}{l|}{16,945}      & 16,300    \\ \cline{2-6} 
                                                     & Clang           & \multicolumn{1}{l|}{19,061}      & \multicolumn{1}{l|}{19,167}      & \multicolumn{1}{l|}{14,513}      & 14,942   \\ \hline
\multirow{2}{*}{ARM32}                               & GCC                                                      & \multicolumn{1}{l|}{19,404}      & \multicolumn{1}{l|}{14,376}            & \multicolumn{1}{l|}{15,201}            &        14,685       \\ \cline{2-6} 
    & Clang           & \multicolumn{1}{l|}{25,349}            & \multicolumn{1}{l|}{16,348}            & \multicolumn{1}{l|}{12,525}            &   13,620     \\ \hline
\multirow{2}{*}{AArch64}    & GCC                                                      & \multicolumn{1}{l|}{22,167}      & \multicolumn{1}{l|}{17,805}      & \multicolumn{1}{l|}{18,780}      & 18,031     \\ \cline{2-6} 
                                                     & Clang        & \multicolumn{1}{l|}{21,720}      & \multicolumn{1}{l|}{20,196}      & \multicolumn{1}{l|}{15,218}      & 15,906   \\ \hline
\end{tabular}
\label{tab_bmerge}
\vspace{-1.0em}
\end{table}


\begin{table}[!thbp]
\caption{The number of equivalent pairs that are generated by the \textsc{BPair} algorithm under 64-bit architectures.}
\renewcommand\arraystretch{1.3}
\centering
\begin{tabular}{|c|c|llll|llll|}
\hline
\multicolumn{1}{|l}{\diagbox[height=22pt,width=54pt]{ISAs}{}} &   \makecell[l]{ISAs\\(Compilers)} & \multicolumn{4}{c|}{\makecell[c]{AArch64\\(GCC)}}                                                                                                        & \multicolumn{4}{c|}{\makecell[c]{AArch64\\(Clang)}}                                                                                                     \\ \cline{2-10} 
                      
    (Compilers)   & Opt levels      & \multicolumn{1}{c|}{\textbf{O0}} & \multicolumn{1}{c|}{\textbf{O1}} & \multicolumn{1}{c|}{\textbf{O2}} & \multicolumn{1}{c|}{\textbf{O3}} & \multicolumn{1}{c|}{\textbf{O0}} & \multicolumn{1}{c|}{\textbf{O1}} & \multicolumn{1}{c|}{\textbf{O2}} & \multicolumn{1}{c|}{\textbf{O3}} \\ \hline
\multirow{4}{*}{\begin{tabular}[c]{@{}c@{}}x86-64\\ (GCC)\end{tabular}}   & \textbf{O0}     & \multicolumn{1}{l|}{35,691}     & \multicolumn{1}{l|}{34,364}     & \multicolumn{1}{l|}{32,632}     & 31,303                         & \multicolumn{1}{l|}{39,760}     & \multicolumn{1}{l|}{44,917}     & \multicolumn{1}{l|}{31,073}     & 30,331         \\ \cline{2-10} 

 & \textbf{O1}     & \multicolumn{1}{l|}{34,702}     & \multicolumn{1}{l|}{42,953}     & \multicolumn{1}{l|}{36,911}     & 39,557            & \multicolumn{1}{l|}{33,844}     & \multicolumn{1}{l|}{33,384}     & \multicolumn{1}{l|}{36,420}     & 35,538  \\ \cline{2-10}

& \textbf{O2}     & \multicolumn{1}{l|}{34,148}     & \multicolumn{1}{l|}{40,884}     & \multicolumn{1}{l|}{39,750}     & 37,134	  & \multicolumn{1}{l|}{32,658}     & \multicolumn{1}{l|}{32,068}     & \multicolumn{1}{l|}{35,188}     & 34,239     \\ \cline{2-10} 

& \textbf{O3}     & \multicolumn{1}{l|}{32,440}     & \multicolumn{1}{l|}{37,926}     & \multicolumn{1}{l|}{37,397}     & 41,367     & \multicolumn{1}{l|}{30,909}     & \multicolumn{1}{l|}{30,279}     & \multicolumn{1}{l|}{35,667}     & 34,841      \\ \hline

\multirow{4}{*}{\begin{tabular}[c]{@{}c@{}}x86-64\\ (Clang)\end{tabular}} & \textbf{O0}     & \multicolumn{1}{l|}{35,961}     & \multicolumn{1}{l|}{32,430}     & \multicolumn{1}{l|}{30,559}  &  29,253  & \multicolumn{1}{l|}{38,640}     & \multicolumn{1}{l|}{44,452}     & \multicolumn{1}{l|}{30,488}     &  29,799     \\ \cline{2-10} 

& \textbf{O1}     & \multicolumn{1}{l|}{44,603}     & \multicolumn{1}{l|}{32,212}     & \multicolumn{1}{l|}{30,187}     & 	28,706 & \multicolumn{1}{l|}{45,092}     & \multicolumn{1}{l|}{45,550}     & \multicolumn{1}{l|}{31,263}     & 30,478   \\ \cline{2-10} 
                                                                          
& \textbf{O2}     & \multicolumn{1}{l|}{30,895}     & \multicolumn{1}{l|}{35,327}     & \multicolumn{1}{l|}{33,538}     & 	33,801  & \multicolumn{1}{l|}{30,973}     & \multicolumn{1}{l|}{31,461}     & \multicolumn{1}{l|}{42,820}     & 41,626      \\ \cline{2-10} 
                                                                          
& \textbf{O3}     & \multicolumn{1}{l|}{30,032}     & \multicolumn{1}{l|}{34,238}     & \multicolumn{1}{l|}{32,646}     & 	33,008 & \multicolumn{1}{l|}{30,084}     & \multicolumn{1}{l|}{30,563}     & \multicolumn{1}{l|}{41,553}     & 41,763     \\ \hline
\end{tabular}
\label{tab_21}
\end{table}

\begin{table}[!htbp]
\vspace{-1.0em}
\caption{The number of equivalent pairs that are generated by the \textsc{BPair} algorithm under 32-bit architectures.}
\renewcommand\arraystretch{1.3}
\centering
\begin{tabular}{|c|c|llll|llll|}
\hline
\multicolumn{1}{|l}{\diagbox[height=22pt,width=54pt]{ISAs}{}} &   \makecell[l]{ISAs\\(Compilers)} & \multicolumn{4}{c|}{\makecell[c]{ARM32\\(GCC)}}                                                                                                        & \multicolumn{4}{c|}{\makecell[c]{ARM32\\(Clang)}}    \\ \cline{2-10} 

(Compilers) & Opt levels      & \multicolumn{1}{c|}{\textbf{O0}} & \multicolumn{1}{c|}{\textbf{O1}} & \multicolumn{1}{c|}{\textbf{O2}} & \multicolumn{1}{c|}{\textbf{O3}} & \multicolumn{1}{c|}{\textbf{O0}} & \multicolumn{1}{c|}{\textbf{O1}} & \multicolumn{1}{c|}{\textbf{O2}} & \multicolumn{1}{c|}{\textbf{O3}} \\ \hline

\multirow{4}{*}{\begin{tabular}[c]{@{}c@{}}x86\\ (GCC)\end{tabular}}   & \textbf{O0}     & \multicolumn{1}{l|}{38,833}     & \multicolumn{1}{l|}{32,740}     & \multicolumn{1}{l|}{31,099}     & 32,639	 & \multicolumn{1}{l|}{40,515}     & \multicolumn{1}{l|}{43,235}     & \multicolumn{1}{l|}{29,914}     & 	28,684        \\ \cline{2-10} 

& \textbf{O1}     & \multicolumn{1}{l|}{33,052}     & \multicolumn{1}{l|}{38,410}     & \multicolumn{1}{l|}{38,124}     & 35,414       & \multicolumn{1}{l|}{31,744	}     & \multicolumn{1}{l|}{30,601}     & \multicolumn{1}{l|}{32,998}     & 31,484                           \\ \cline{2-10}

& \textbf{O2}     & \multicolumn{1}{l|}{31,474}     & \multicolumn{1}{l|}{35,606}     & \multicolumn{1}{l|}{36,512}     & 34,443	 & \multicolumn{1}{l|}{29,950}     & \multicolumn{1}{l|}{28,413}     & \multicolumn{1}{l|}{31,526}     & 30,217          \\ \cline{2-10} 

 & \textbf{O3}     & \multicolumn{1}{l|}{30,310}     & \multicolumn{1}{l|}{33,167}     & \multicolumn{1}{l|}{33,845}     & 37,675  & \multicolumn{1}{l|}{28,733}     & \multicolumn{1}{l|}{27,306}     & \multicolumn{1}{l|}{31,860}     & 30,585        \\ \hline

\multirow{4}{*}{\begin{tabular}[c]{@{}c@{}}x86\\ (Clang)\end{tabular}} & \textbf{O0}     & \multicolumn{1}{l|}{37,095}     & \multicolumn{1}{l|}{31,599}     & \multicolumn{1}{l|}{31,686}     & 30,161    & \multicolumn{1}{l|}{37,773}     & \multicolumn{1}{l|}{44,006}     & \multicolumn{1}{l|}{30,506}     & 29,446         \\ \cline{2-10} 

& \textbf{O1}     & \multicolumn{1}{l|}{44,589}     & \multicolumn{1}{l|}{31,719}     & \multicolumn{1}{l|}{30,935}     & 29,336     & \multicolumn{1}{l|}{44,758}     & \multicolumn{1}{l|}{44,195}     & \multicolumn{1}{l|}{30,686}     & 	29,438   \\ \cline{2-10}

& \textbf{O2}     & \multicolumn{1}{l|}{30,995}     & \multicolumn{1}{l|}{34,734}     & \multicolumn{1}{l|}{34,584}     & 35,203 & \multicolumn{1}{l|}{31,029}     & \multicolumn{1}{l|}{30,613}     & \multicolumn{1}{l|}{42,247}     & 40,385     \\ \cline{2-10}

& \textbf{O3}     & \multicolumn{1}{l|}{30,227}     & \multicolumn{1}{l|}{33,785}     & \multicolumn{1}{l|}{33,581}     & 34,291   & \multicolumn{1}{l|}{30,281}     & \multicolumn{1}{l|}{29,697}     & \multicolumn{1}{l|}{40,825}    & 40,469      \\ \hline
\end{tabular}
\label{tab_22}
\end{table}

For the same function coming from different architectures, optimization levels or compilers, we can efficiently extract equivalent basic blocks pairs by invoking the~\textsc{BPair} algorithm. Table~\ref{tab_21} and Table~\ref{tab_22} show the resulting equivalent assembly code pairs targeting 64-bit and 32-bit platforms, respectively. It's worth noting that all resulting assembly pairs shown in Table~\ref{tab_21} and Table~\ref{tab_22} have been normalized and all duplicates have also been eliminated.

\subsection{Application in Similarity Detection}

The important application of \textsc{BinSimDB} is that it can be used to train machine learning models capable of detecting fine-grained similarities between binary code snippets.
To illustrate this point, we attempt different data combinations from the proposed dataset, and train the Transformer-based detector introduced in \ref{sec:classifier}.
It should be clarified that, the basic blocks that have more than 100 instructions will be truncated when training the machine learning model, because we found only 0.45\% of cases in our dataset fall into this category. Please note that what we do herein will also lay the groundwork for the subsequent experiments in Section 4.3 and Section 4.4.

We first extract datasets towards 32-bit platforms, denoted as $\mathcal{D}\text{-}\texttt{x86(GCC)/}$
$\texttt{ARM32(GCC)}$. In total, there are 750,000 basic block pairs in this dataset, where half are equivalent pairs while the other half are not. We randomly choose 80\% instances as the training set and the remaining 20\% as the testing set. Two basic blocks in a pair are across different optimization levels, and also different architectures, i.e., x86 and ARM32. Additionally, GCC is leveraged as a control variable, so all binaries in this dataset are built using the same compiler. Besides, we re-use the method proposed in~\cite{zuo2019neural} to produce nonequivalent basic block pairs. The evaluation result shows a well-trained model can finally achieve an AUC value as high as 99.4\% over the testing set. The AUC (Area Under the Curve) value is a scale-invariant performance measurement for classification models. Specifically, it is used with Receiver Operating Characteristic (ROC) curves. A higher AUC value indicates a better measure of separability, that is the model effectively differentiates between the positive and negative classes.

\begin{figure*}[!htbp]
\centering
\subfloat[Cross-ISAs \& opt-levels]{\includegraphics[scale=0.32]{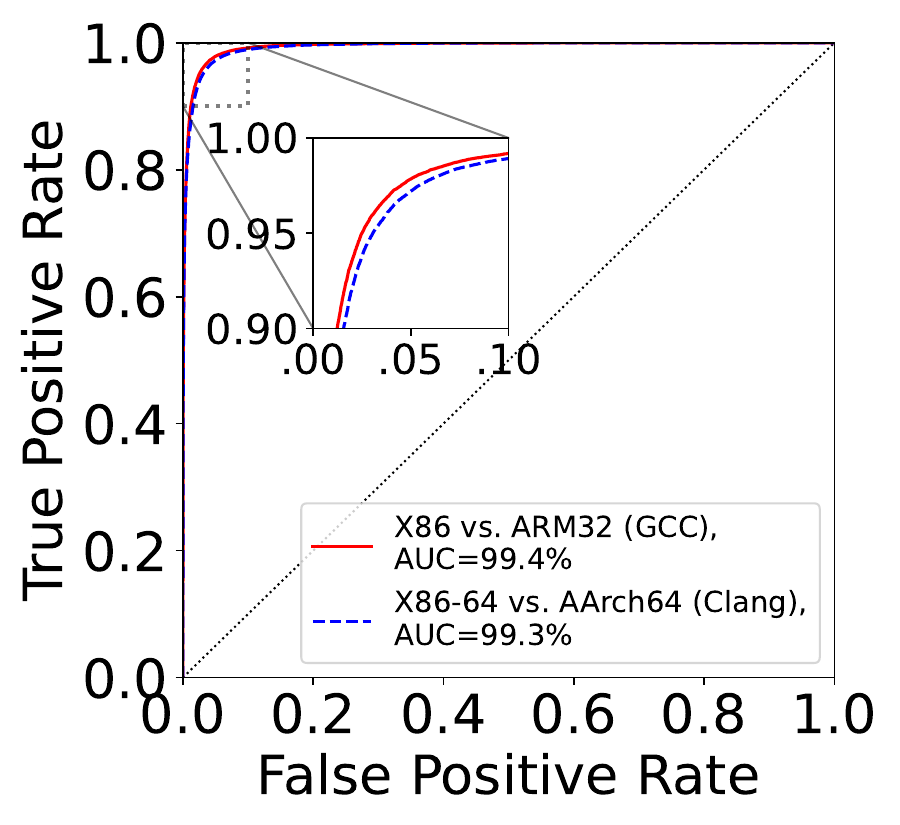}}\quad\quad
\subfloat[Cross-ISAs, opt-levels \& compilers]{\includegraphics[scale=0.32]{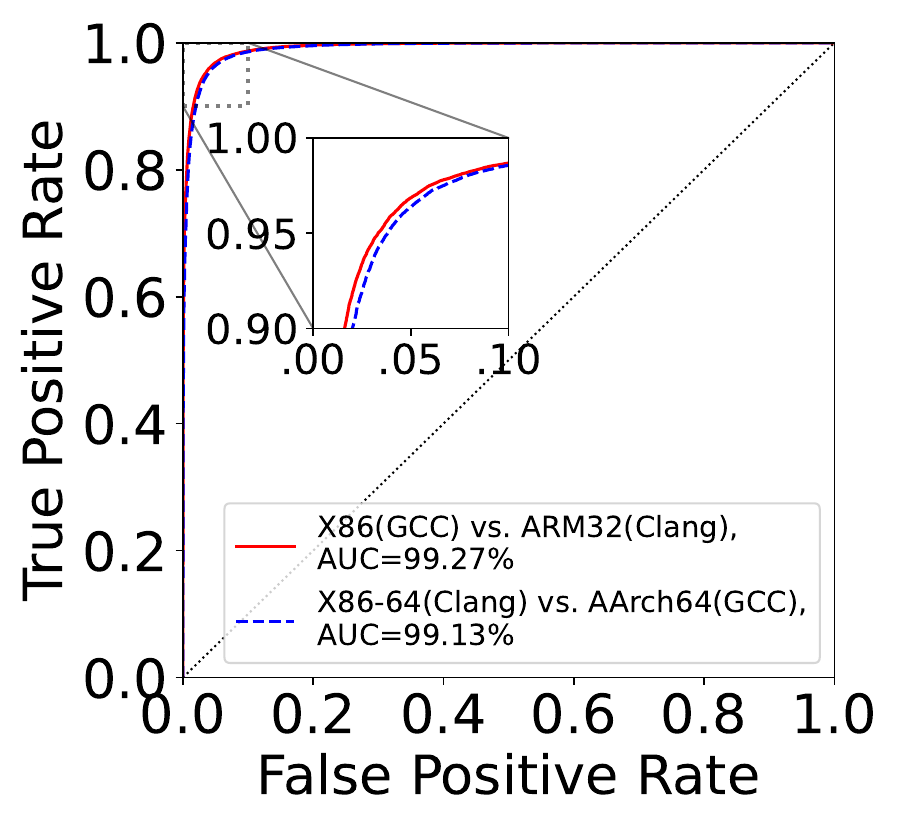}}\\
\subfloat[x86 vs. ARM32]{\includegraphics[scale=0.34]{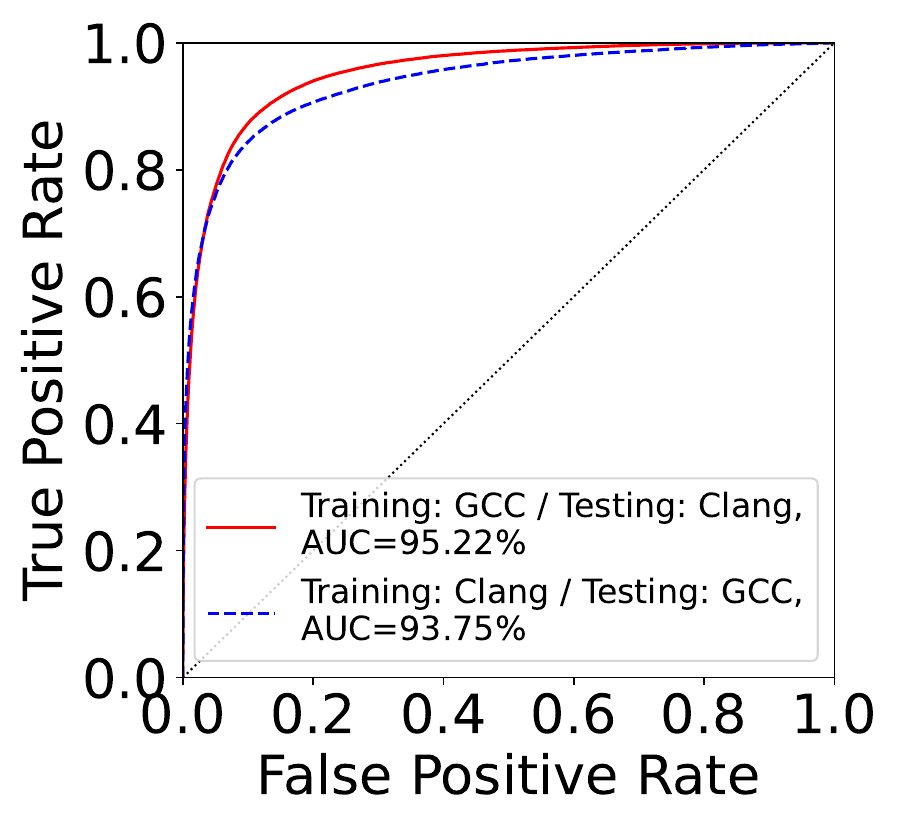}}\quad\quad
\subfloat[x86-64 vs. AArch64]{\includegraphics[scale=0.34]{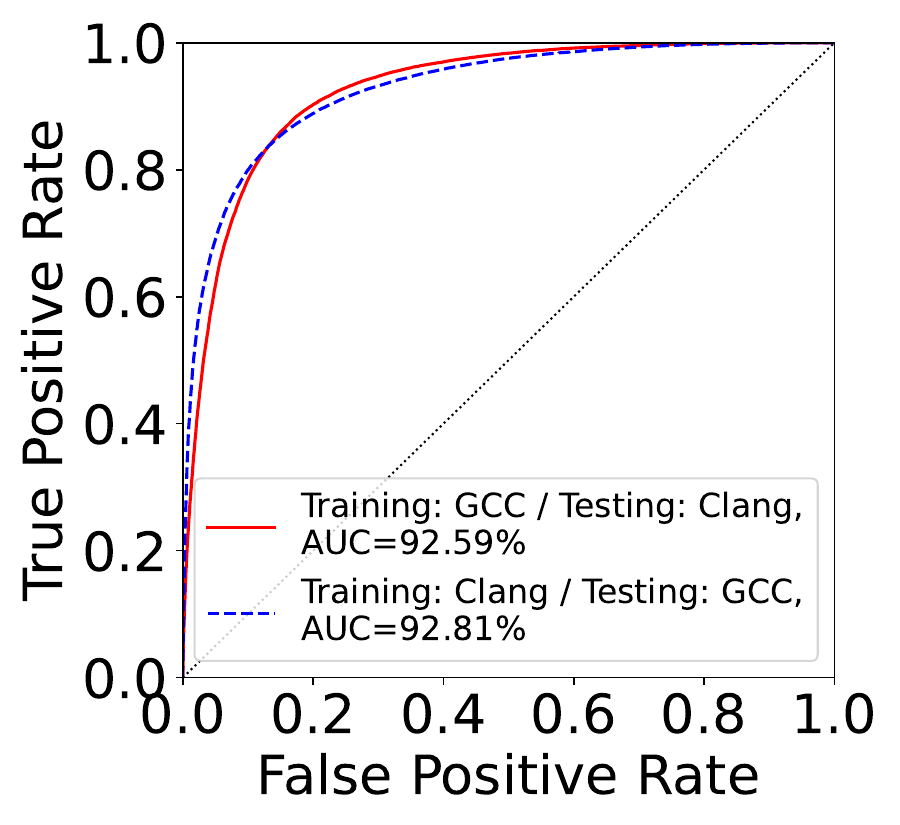}}
\caption{Similarity detection among basic blocks based on \textsc{BinSimDB}. }\label{fig:eval_roc12}
\end{figure*}

Following the similar setting, we can obtain another dataset, which is towards 64-bit platforms, namely $\mathcal{D}\text{-}\texttt{X86}\text{-}\texttt{64(Clang)/AArch64(Clang)}$. Herein, the different aspect is two basic blocks in a pair are from x86-64 and AArch64 architectures, respectively. Plus, all binaries in the dataset are built using Clang. The evaluation result shows the model can achieve an AUC value as 99.3\% over the testing set. The ROC curve is shown in Figure~\ref{fig:eval_roc12}(a). 

We further investigate the performance of models targeting cross-compiler challenge. To this end, we consider a dataset, $\mathcal{D}\text{-}\texttt{x86(GCC)/ARM32(Clang)}$. The size and distribution proportion of this dataset follow the same settings as aforementioned. However, two basic blocks in a pair not only come from different architectures, but also are obtained by different compilers. Namely, x86 and ARM32 binaries are built with GCC and Clang, respectively. The evaluation result shows the model can achieve a high AUC value as 99.27\% over the testing set. Based on another dataset focusing on 64-bit platforms, 
$\mathcal{D}\text{-}\texttt{x86}\text{-}\texttt{64(Clang)/}$
$\texttt{AArch64(GCC)}$, we also can observe an analogous performance, as Figure~\ref{fig:eval_roc12}(b) shows. All in all, the performance of our models remains stable no matter which compilers or architectures are involved, proving that the models manifest better generalization ability. 

\begin{figure}[!hbp]
\centering
\subfloat[x86(GCC) vs. ARM32(GCC)]{\includegraphics[scale=0.35]{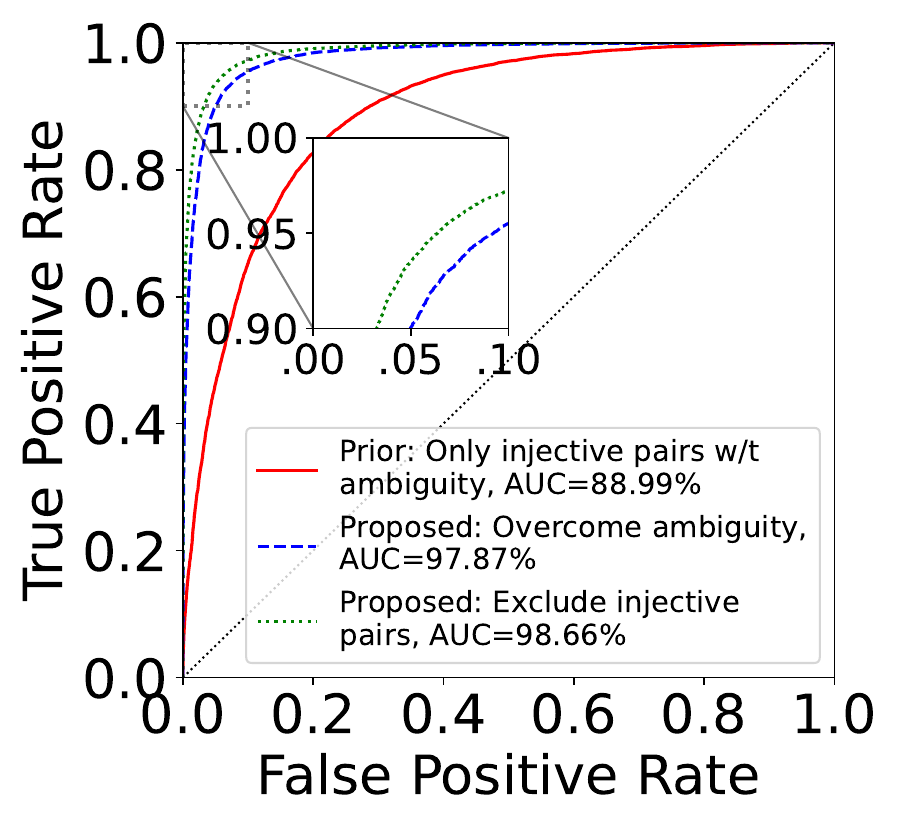}}\quad\quad
\subfloat[x86-64(Clang) vs. AArch64(GCC)]{\includegraphics[scale=0.35]{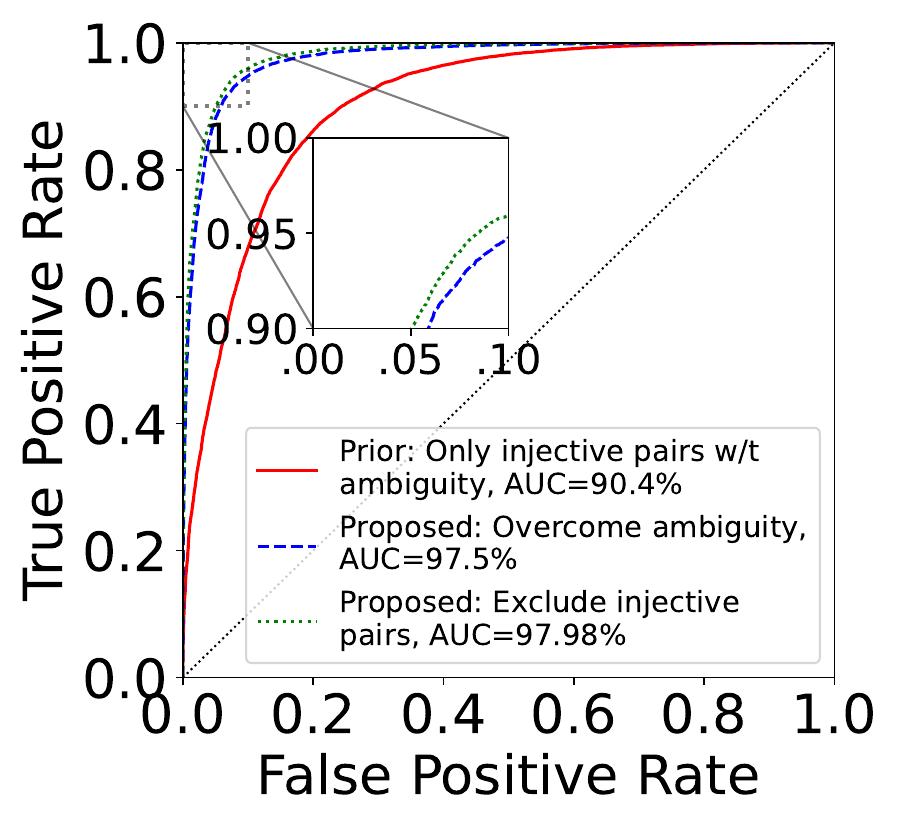}}
\caption{The advantage of \textsc{BinSimDB} over prior method\cite{zuo2019neural}. }\label{fig:eval_roc34}
\end{figure}

\vspace{3pt}
\noindent\textbf{Transferability.} Different compilers will likely end up emitting noticeably distinct assembly code, though the obtained programs should still behave in exactly the same way. The discrepancy among compilers usually poses challenge in discovering semantic equivalence between binaries especially in the cross-architecture scenario. Therefore, we are interested in understanding the extent of transferability of a learning-based model can demonstrate when faced with previously unseen samples generated by a different compiler. For this purpose, we first construct two dis-joint datasets towards 32-bit platforms, $\mathcal{D}_{Train}\text{-}\texttt{x86/ARM32}$ $\texttt{(GCC)}$ and $\mathcal{D}_{Test}\text{-}\texttt{x86/ARM32(Clang)}$. The size and distribution proportion of datasets follow the same settings as aforementioned. The change is that all binaries in the training set are compiled by GCC. In contrast, all binaries in the testing set are compiled by Clang. The evaluation result shows the model can achieve an AUC value of 95.22\% over the testing set. If the binaries in the training set are compiled by Clang, while the binaries in the testing set are compiled by GCC. We can observe a very close AUC value of 93.75\%, as Figure~\ref{fig:eval_roc12}(c) shows.

We additionally conduct a complementary study to assess the extent to which this compiler-agnostic character can be maintained by a learning-based model towards binaries on 64-bit platforms. We first involve two dis-joint datasets, that is $\mathcal{D}_{Train}\text{-}\texttt{x86}\text{-}\texttt{64}\texttt{/AARCH64}$$\texttt{(GCC)}$ and $\mathcal{D}_{Test}\text{-}\texttt{x86}\text{-}\texttt{64/AARCH64(Clang)}$. In other words, all binaries in the training set are compiled by GCC. However, the binaries in the testing set are compiled by Clang. The evaluation result shows the model can achieve an AUC value of 92.59\% over the testing set. If we swap the compilers for the training and testing sets, the AUC value remains relatively stable, i.e., 92.81\%, as Figure~\ref{fig:eval_roc12}(d) shows.

\subsection{Comparison Study}\label{sec:comp}

In the previous study such as~\cite{zuo2019neural}, a pair of semantically equivalent basic blocks under distinct architectures across different optimization levels rely on simplified evidence to build a connection, that is the two basic blocks exclusively come from certain lines of source code and such source code are not used to generate any other basic blocks. If there is any ambiguity that cannot be straightforwardly handled, the basic blocks will be directly abandoned. Hence, a large number of semantically intense code gadgets may ultimately be excluded from the resulting dataset, such as the ternary operator expression shown in Figure~\ref{fig:patch}. 

To verify our insights, we extract such complicated cases to form a testing set towards 32-bit platforms, consisting of 43,000 basic block pairs. As Figure~\ref{fig:eval_roc34}(a) shows, if we adopt the approach~\cite{zuo2019neural} to construct a training set, the classifier trained with this dataset can achieve an AUC value as 88.99\% on the testing set. If we use the proposed \textsc{BMerge} and \textsc{BPair} to overcome the ambiguity, thus obtained training set can evidently improve the AUC value of the classifier to 97.87\%. If the injective mapping between two cross-architecture basic blocks can be easily inferred, such a basic block pair is considered as a simple case. Even though we further exclude those simple cases from training set, the performance of trained model remains stable, with an AUC value of 98.66\%. The size of all training sets is 172,000, so that the testing set takes up 20\% of the overall data. 

When considering 64-bit platforms, and cross-compiler scenarios, we can observe a similar result. As Figure~\ref{fig:eval_roc34}(b) shows, by using \textsc{BMerge} and \textsc{BPair} to construct the training set, we can achieve an AUC value at 97.88\% and 97.5\% depending on whether the simple cases are removed or not. This is obviously higher than the comparative group with an AUC value of 90.4\%. Therefore, we can conclude that the proposed dataset construction approach can greatly help to improve the performance of binary code similarity comparison.

\begin{figure*}[!htbp]
\centering
\subfloat[x86-64 (GCC)]{\includegraphics[scale=0.36]{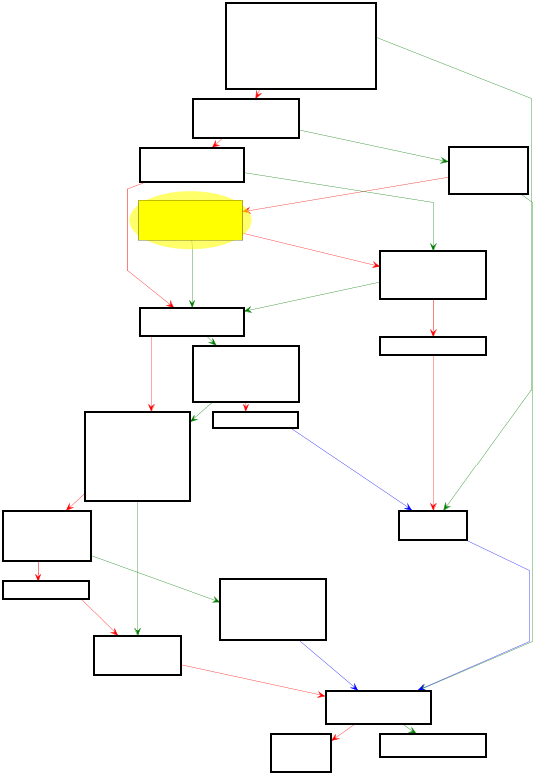}}\quad\quad 
\subfloat[AArch64 (Clang)]{\includegraphics[scale=0.36]{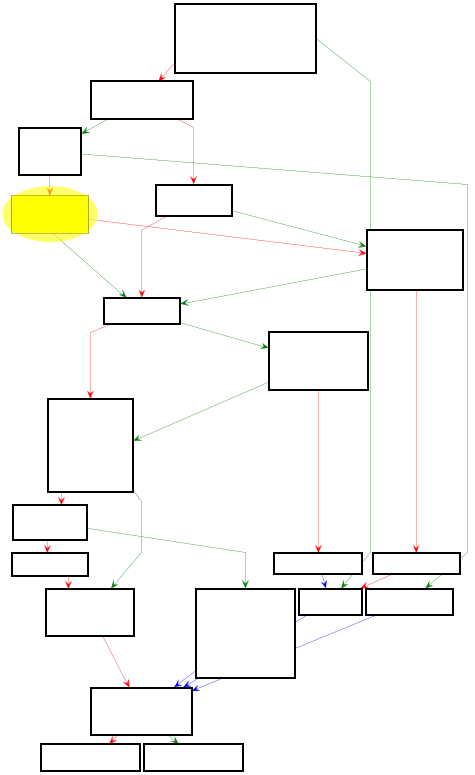}}\\ 
\subfloat[AArch64 (Clang) with obfuscation]{\includegraphics[scale=0.35]{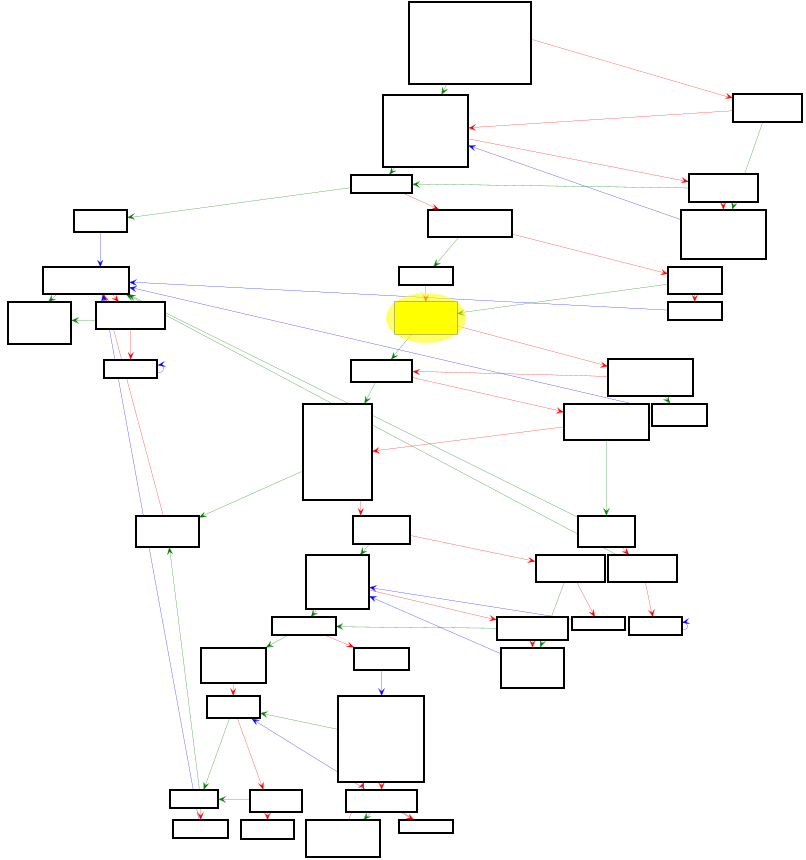}}
\caption{The CFGs of \texttt{tftp\_connect} function in \texttt{libcurl}. }\label{fig:showcase}
\end{figure*}

\subsection{Case Study}

Again, the focus of this work is on the construction of the dataset, rather than developing new advanced BCSA techniques. To demonstrate the application of the \textit{fine-grained} binary code similarity comparison in real-world scenarios, 
we conducted an experiment for the purpose of patch detection. CVE-2019-5482 is a heap buffer overflow vulnerability in the TFTP protocol handler of \texttt{libcurl}. The affected versions range from 7.19.4 to 7.65.3. 
Figure \ref{fig:showcase}(a) shows the CFG of \texttt{tftp\_connect} function from the upgraded version 7.66.0, where the patched basic block is highlighted in color. The binary is built in GCC towards an x86-64 platform, and the default optimization level is adopted. Locating the patch in a function with hundreds of basic blocks is a time-consuming task, but it is crucial for security engineers. The existing function-level BCSA approaches have difficulties in solving this kind of problems. Figure \ref{fig:showcase}(b) shows the CFG of a patched function but from the version 7.67.0. We build the binary towards an AArch64 platform using Clang as the compiler. Our experiment shows that a detector introduced in Section~\ref{sec:classifier}, trained using the proposed fine-grained dataset, is able to quickly and accurately identify the affected basic block, namely the highlighted one. Moreover, we leverage the \textit{bogus control flow insertion}\cite{junod2015obfuscator} to generate an obfuscated binary, as Figure \ref{fig:showcase}(c) shows. Still, our method can rapidly locate the corresponding basic block. This definitely could save a significant amount of time in security analysis. By contrast, we cannot find well-matched basic blocks in an earlier version such as 7.65.0 because it is unpatched. All the results are manually verified.

\section{Related Work}\label{sec:relate}


\subsection{Binary Similarity Analysis}

The advances in applied deep learning, such as graph learning and natural language processing (NLP), have inspired many new methods for comparing the binary code similarity. Given a large body of research in the pertinent area, our literature review is not intended to be exhaustive.

\vspace{3pt}
\noindent\textbf{Code Learning.} Assembly code can be considered as a sequence of tokens, therefore numerous researchers shift their focus to the arsenal of NLP techniques when tackling the binary function similarity problem, for example, \texttt{BinDNN}~\cite{lageman2017bin},
\texttt{Asm2Vec}~\cite{ding2019asm2vec},
and \texttt{SAFE}~\cite{massarelli2019safe}, etc.
Later, Yu et al. utilized the BERT model pre-trained on four tasks to learn semantics of assembly code~\cite{yu2020order}. Similarly, \texttt{BinShot}~\cite{ahn2022practical} is also based on BERT model to detect binary code similarity. In addition, \texttt{jTrans}~\cite{wang2022jtrans} proposed a binary function embedding method using the jump-aware Transformer-based model. However, all of these approaches detect similarity between functions, and cannot tackle the case when only partial code within a function is under consideration. 

\vspace{3pt}
\noindent\textbf{Graph Learning.} 
There are a surge of works on binary code similarity analysis, which adopt existing graph based techniques. For example, \texttt{Structure2Vec}~\cite{dai2016discriminative} is adapted by Xu et al. to learn CFG embeddings ~\cite{xu2017neural}. 
Besides, Yu et al.~\cite{yu2020order} use an adjacency matrix to represent a function CFG, then a CNN is further applied to generate embeddings. However, these methods suffer from a failure to capture the rich contextual information in assembly code (e.g., intra- or inter-procedural control flows). To address this problem, \texttt{DeepBinDiff}~\cite{duan2020deepbindiff} builds inter-procedural CFG (ICFG) based on the call graph and CFGs to provide program-wide contextual information. In particular, Text-associated DeepWalk algorithm (TADW)~\cite{yang2015network} is used to learn vector representation for each vertex in a graph. Furthermore, Kim et al.~\cite{kim2022improving} leverage the graph convolutional networks based graph alignment technique~\cite{wang2018cross} to learn contextual information. However, this method relies on partial cross-platform alignment information as a priori.


\subsection{Dataset Construction}

Unlike computer vision and natural language which have a large body of well-labelled data, high-quality datasets are a kind of precious or even scarce resource in cybersecurity. 
Many researchers consider the unavailability of well maintained data is a common barrier in the area of cybersecurity~\cite{tian2012identifying,wang2021patchdb,zuo2024vulnerability}. Therefore, we have seen a few recent works focusing on dataset construction, for example, system provenance dataset \texttt{ProvSec}~\cite{shrestha2023provsec}, and source code patch dataset \texttt{PatchDB}~\cite{wang2021patchdb}. However, the datasets targeting binary analysis are still a remaining issue which needs to be addressed by the security community. While some groups open-sourced their datasets, the contamination or degradation of data quality has been observed. For example, previous work~\cite{andriesse2017compiler} studied the dataset~\cite{bao2014byteweight}, which is used for the function boundaries detection in binary code, and ``\textit{found many functions to be duplicated across the training and testing sets, thus artiﬁcially increasing their F1-score}''~\cite{qiao2017function}. In addition, the datasets for binary code similarity research are very rare. After investigating 43 papers in this area, Kim et al.~\cite{kim2022revisiting} found ``\textit{only two of them opened their entire dataset}'', which leads to be often infeasible for reproducing the previous results. In our previous research, we also encountered similar issues. For example, \texttt{CrossMal}~\cite{song2022inter} is a dataset consisting of cross-architecture function pairs 
based on IoT binaries. 
The download link provided in the paper had became inaccessible when we wrote this article. 

A few new datasets for the binary similarity analysis task have been proposed in recent years\cite{marcelli2022machine, zuo2023comp, kim2022revisiting}.
However, these datasets cannot support a fine-grained study such as the similarity comparison at a basic block level. By contrast, Zuo et al.~\cite{zuo2019neural} propose an approach, which can automatically generate equivalent basic block pairs. But, their method extracts binary code directly from the backend of compilers, which still has a certain gap from practical application because the binary code obtained from decompilation is not exactly identical to the binary code generated by the compiler.

\subsection{Usage Cases of Binary Similarity}

Binary similarity has been used for multiple usage cases. Vulnerability detection and patch detection is one of well-known application cases. We presented a case study to detect a vulnerability and a patch in our paper in the evaluation section. Another popular application is the detection and correlation of malicious software which are found in various fields across Enterprise \cite{10.1145/2660267.2660330,10.1007/978-3-030-30215-3_18}, IoT \cite{10.1145/3427228.3427256,NGO2020280}, and Energy fields \cite{8366503,defcon24}. 
To reach out to a large scope of targets, there are even types of malware that work across platforms (e.g., OS) and architectures. Binary similarity techniques will be a useful foundational technique to extend the applicability of the defense techniques.

\section{Discussion and Future Work}\label{sec:discuss}

The main purpose of \textsc{BinSimDB} is to facilitate BCSA research. However, because we have established a strong connection between assembly code and corresponding source information as the ground truth, the current dataset can be easily extended for diverse research purposes. For example, we can further build up a dataset containing semantically meaningful code gadgets, such as a loop structure. This dataset can be used to investigate the loop detection problem in binary analysis. We consider this as an interesting exploration direction.

Large Language Models (LLMs) have achieved remarkable success in various natural language processing tasks. As we have developed automated scripts that can continuously generate large volumes of high-quality data, this also paves the way for binary analysis research based on LLMs. We believe the intersection of LLMs and binary analysis has the potential to inspire further advancements in relevant areas, and accordingly plan to conduct more related study in the future.


\section{Conclusion}\label{sec:conclude}

This paper presents the construction of~\textsc{BinSimDB}, a 
fine-grained dataset for research on binary code similarity analysis. To maintain the high quality of data, we extract ground truth from source code information. We propose two algorithms, \textsc{BMerge} and \textsc{BPair}, specifically designed to address the challenges of matching semantically equivalent binary code snippet pairs across different architectures and optimization levels. The proposed algorithms can preserve a good coverage and provide fine granularity to the greatest degree. Furthermore, we conduct comprehensive experiments to investigate the properties of the constructed dataset and demonstrate the strength of our proposed algorithms. The evaluation results show that~\textsc{BinSimDB} is promising to facilitate the binary code similarity analysis.

\section*{Acknowledgement}


Sandia National Laboratories is a multimission laboratory managed and operated by National Technology and Engineering Solutions of Sandia, LLC., a wholly owned subsidiary of Honeywell International, Inc., for the U.S. Department of Energy’s National Nuclear Security Administration under contract DE-NA0003525. This article describes objective technical results and analysis. Any subjective views or opinions that might be expressed in the article do not necessarily represent the views of the U.S. Department of Energy or the United States Government.
This work was supported through contract CR-100043-23-51577 with the U.S. Department of Energy.

%
%
%
\bibliographystyle{splncs04}

\end{document}